\documentclass[runningheads,citeauthoryear]{apinv}
\usepackage{epsfig,cite}
\usepackage[T2A]{fontenc}
\usepackage[cp1251]{inputenc}
\usepackage{amsmath,amssymb}
\usepackage{multirow}
\usepackage{graphicx}
\usepackage{epsfig}
\usepackage{subfigure}
\usepackage[english]{babel}
\usepackage{url}
\RequirePackage{color}
\usepackage{ifpdf}
\usepackage{epstopdf}
\newcommand{\Msol}{M_{ \odot}}
\begin{document}

\title{Numerical stability of the electromagnetic quasinormal and quasibound modes of Kerr black holes}
\titlerunning{Stability of the electromagnetic QNM and QBM of Kerr BH}
\author{Denitsa Staicova\inst{1}, Plamen Fiziev\inst{2,3}}
\authorrunning{D. Staicova, P. Fiziev}
\tocauthor{Denitsa Staicova, Plamen Fiziev}
% Command tocautor{} is used by the Latex to give author names
% to the Contents of the volume (automatically generated)
\institute{Institute for Nuclear Research and Nuclear Energy,
Bulgarian Academy of Sciences, Bulgaria
	\and Sofia University Foundation for Theoretical and Computational Physics and
Astrophysics
	\and JINR, Dubna, Russia  \newline
	\email{dstaicova@inrne.bas.bg}  \newline
	\email{fiziev@phys.uni-sofia.bg} }
\papertype{Submitted on xx.xx.xxxx; Accepted on xx.xx.xxxx}	
% Papertype can be "Research report", "Review", "Invited lecture", "Conference talk",
% "Conference poster", "Lecture at scientific seminar", "Summary of dissertation",  etc.
\maketitle

\begin{abstract}
The proper understanding of the electromagnetic counterpart of gravity-waves emitters is of serious interest to the multimessenger astronomy. In this article, we study the numerical stability of the quasinormal modes (QNM) and quasibound modes (QBM) obtained as solutions of the Teukolsky Angular Equation and the Teukolsky Radial Equation with appropriate boundary conditions. We use the epsilon-method for the system featuring the confluent Heun functions to study the stability of the spectra with respect to changes in the radial variable. We find that the QNM and QBM are stable in certain regions of the complex plane, just as expected, while the third ``spurious'' spectrum was found to be numerically unstable and thus unphysical. This analysis shows the importance of understanding the numerical results in the framework of the theory of the confluent Heun functions, in order to be able to distinguish the physical spectra from the numerical artifacts.
\end{abstract}

\section{The spectra of the black holes and other compact massive objects in the Kerr space-times.}

The quest for understanding ultra-energetic events such as gamma-ray bursts, active-galactic nuclei, quasars etc. passes through understanding of the compact massive objects which are thought to be their central engine -- the rotating black holes, magnetars, binary neutron stars, etc.. For this, one needs to make a clear difference between theoretical, numerical and observational black hole (BH) and also, to keep account of the strengths and the weaknesses of the methods -- numerical and observational -- when working with the different  types of objects.

In the case of black holes candidates in the role of central engines, the theoretical energy extraction processes of ultra-relativistic jets (such as the Penrose process or the Blandford-Znajek process) should depend on the rotational BH parameter ($a_*$), with a predicted threshold for effective jet launching to be around $a_*>0.5$, with more conservative estimate at $a_*\sim 0.8$ \cite{AGN}. Observationally, the measured rotational parameters seem to follow this requirement -- a list of measured $a_*$ can be found in Table 1 in \cite{BHrot}. From it, one can see that in general, the measurements correspond to the expectations. There is, however, a catch. The measurement of the spins of the black hole candidates is not independent -- it depends critically on the measurement of their masses, which comes with serious incertitude \cite{rot}. One such example is the black hole candidate GRO J0422+32. Its mass was estimated to $3.97\pm 0.95 \Msol$ \cite{GRO1} but further analysis showed that its mass can be as low as $2.1\Msol$ \cite{GRO2} -- thus putting it in the mass-interval of neutron stars ($M\le 2.2\Msol$)). In the same article, a number of other black hole candidates have their masses re-examined and corrected, which poses a serious question about	 all the results on black hole candidates. Additionally, in \cite{super-spinars,super-spinars2} it was demonstrated that numerous black hole candidates can be fitted by super-spinars (objects with $a_*>1$), thus showing that
the X-ray spectrum is not enough to fix the spin of the black hole candidate.  Considering the importance of the mass-spin measurement for the modeling of astrophysical jets, it is clear that there is an acute need for a better understanding of the theory of the rotating black hole.

The vacuum solution of the Einstein equations which may be used for description of a rotating massive compact object (black hole or super-spinar) is the Kerr metric. The Kerr metric depends only on 2 parameters -- the rotation $a$ \footnote{ We will use here $a$ to denote the theoretical rotational parameter, while the observational will be $a_*$. Usually the mass of the Kerr metric is taken for a constant so $a_*=a/M$.} and the mass $M$. While there is a known theoretical problem in using the Kerr metric for objects different from rotating black holes (the problem of finding interior ''matter`` solution matching the exterior vacuum Kerr solution), it is thought that it approximates well enough any distant rotating compact massive object, the precise type of the object being specified by the boundary conditions imposed additionally on the given by the metric spacetime to fix its physical content, like black holes, super-spinars, etc. . Confirming that the observational spectra indeed coincide with the Kerr {\em black hole} spectra, is an important task facing astronomy and astrophysics.

For a long time, the theory of perturbations was the only viable method of dealing with different type of matter in the Kerr metric, or to study evolution of several interacting black holes. In the past decade, the perturbation theory has been replaced by the much more sophisticated, but also computationally expensive and complicated numerical relativity simulations method. The theory of perturbations, however, still has place in the study of this type of objects, allowing a clearer view of the purely gravitational properties of the system. Moreover, the important role of perturbative approach has been confirmed by full-GR numerical calculations. In the theory of perturbations, the characteristic ring-down of the black hole, as a result of perturbation, is dominating in later epochs through the complex frequencies called the quasi-normal modes (QNMs). The QNMs were used for the first wave-templates for gravitational detectors. They were also observed in the late-time evolution of GRMHD simulations. Finally, QNMs were used in full GR simulations as a marker of the end of the merger phase \cite{QNMsim}.

In \cite{arxiv2014} we presented the quasinormal and quasibound spectra of the Kerr metric obtained by using a novel approach of solving the governing equations in terms of the confluent Heun functions and imposing the required boundary conditions directly on them.  While this method allowed us to obtain both well-known old and also some new results, the novelty of the method makes its numerical stability uncharted.  In the current paper, we study in detail the numerical stability of those spectra which is crucial for further use of the method and the results. In particular, we demonstrate the numerical stability with respect to deviations in the actual radial infinity and the position in the complex plane of the radial variable, proving that the so obtained quasinormal and quasibound spectra are indeed stable, as expected from the theory, and we show a counter-example of a spurious spectrum which despite being a root of the system of equations under consideration, does not satisfy the criteria for numerical stability and is thus non-physical.

\section{The Teukolsky Angular and Radial Equations}
The linearized perturbations of the Kerr metric are described by the Teukolsky Master equation, from which under the substitution $\Psi=e^{i(\omega t+m\phi)}S(\theta)R(r)$ one obtains the Teukolsky Radial Equation (TRE) and the Teukolsky Angular Equation (TAE) \cite{Teukolsky, Teukolsky0,Teukolsky-Kerr} ( where $m=0, \pm 1, \pm 2$ for integer spins and $\omega$ is the complex frequency). The TAE and TRE  describe on the same footing perturbations with different spin ( scalar waves, fermions, electromagnetic and gravitational waves) without taking into account of the origin of the perturbation. Their numerical solutions have provoked the creation of numerical methods aimed at solving the non-trivial transcendental system of differential equations through different approximations, the most well-known being the method of continued fractions and that of direct integration performed by Leaver and Andresson (\cite{QNM, det1,QNM2, Leaver, Leaver0, Andersson, special1, Fiziev1}, \\ \cite{QNM1, AS1, special3, Fiziev3}. For a review on the methods for studying the QNM spectra, see \\ \cite{QNM21}.

The TRE and the TAE are coupled differential equations for whose solutions one needs to impose the appropriate boundary conditions on the boundaries of the physical interval of interest. Depending on the boundary conditions, one obtains as solutions of system the quasinormal spectra (black hole boundary conditions, see below), the quasibound spectra (resonant waves) or total-transmission modes (waves moving in one direction, i.e. entirely to the left or to the right). All of them are discrete spectra of complex, possibly infinite number of frequencies (numbered by $n$), which dominate the late-time evolution of the object's response to perturbations. Unlike normal modes, QNMs do not form a complete set, because the mathematical problem is not self-adjoint. Therefore, one cannot analyze the behavior of the system entirely in terms of QNMs. The QNMs, however, can be used to discern a black hole from other compact massive object -- such observational test was proposed in \cite{QNMtest}, and for a theoretical estimate of the EM output for non-rotating BH -- \cite{QNMtest1}.

Explicitly, the TAE and the TRE are:
\begin{footnotesize}
 \begin{multline}
 \Big(\left(1\!-\!u^2\right)S_{lm,u}\Big)_{,u}+
\left((a\omega u)^2+2a\omega
su\!+\!{}_sE_{lm}\!-\!s^2-\frac{(m\!+\!su)^2}{1\!-\!u^2}\right)S_{lm}=0,
\label{TAE}
\end{multline}
\end{footnotesize}
and
\begin{footnotesize}
\begin{multline}
{\frac {d^{2}R_{\omega,E,m}}{d{r}^{2}}} + (1+s)  \left( {\frac{1}{r-{\it r_{+}}}}+
{\frac{1}{r-{\it r_{-}}}} \right){\frac {dR_{\omega,E,m}}{dr}} +                                      
+\Biggl( {\frac { K ^{2}}{ \left( r-r_{+} \right) \left( r-r_{-} \right) }}- \\
is \left( {\frac{1}{r-{\it r_{+}}}}+ {\frac{1}{r-{\it r_{-}}}} \right)  K- 
-\lambda - 4\,i s \omega r \Biggr)
{\frac {R_{\omega,E,m}} {( r-r_{+})( r-r_{-})}}=0
\label{TRE}
\end{multline}
\end{footnotesize}
 \noindent where $\Delta=r^2-2Mr+a^2=(r-r{_-})(r-r{_+})$, $K=-\omega(r^2+a^2)-ma$,
$\lambda=E-s(s+1)+a^2\omega^2+2am\omega$ and $u=\cos(\theta)$. We fix $s=-1$ for the spin of the perturbation and $r_{\pm}=M\pm \sqrt{M^2-a^2}$ are the two horizons of the BH.

The problem has two known parameter: the mass $M$ and the rotation $a$ and two unknown parameters: the frequency $\omega$ and the constant of separation of the angular equation $E$. Depending on the sign of the imaginary part of the frequency  one can consider both perturbations damping with time, $\Im(\omega)>0$, (the QNM case) and also perturbations growing exponentially with time, $\Im(\omega)<0$ (the QBM case).

For details on the different classes of solutions of the TAE and the TRE EM perturbations ($s=-1$), see \cite{Fiziev3, arxiv2014}. Here we will present only the equations we use to solve the system.

{\bf The equation governing the spectrum of the TAE is}:
$$S_{1,2}(\theta)=e^{\alpha_{1,2}z_{1,2}} z_{1,2}^{\beta_{1,2}/2} z_{2,1}^{\gamma_{1,2}/2}{\text{HeunC}(\alpha_{1,2},\beta_{1,2},\gamma_{1,2},\delta_{1,2},\eta_{1,2}, z_{1,2})}$$

\noindent where $z_{1}=\cos(\theta/2)^2$, $z_{2}=\sin(\theta/2)^2$ and the parameters are:

For the case $m=0$:
$\alpha_1= -\alpha_2 = 4\,a\omega,
\beta_1 = \beta_2 = 1,
\gamma_1 = -\gamma_2 = - 1,
\delta_1 = -\delta_2 = 4\,a\omega,
\eta_1 (\omega)= \eta_2 (-\omega) = 1/2-E-2\,a\omega-{a}^{2}{\omega}^{2}$ and

For the case $m=1$:
$\alpha_1=\alpha_2=-4\,a\omega,
\beta_1=\gamma_2 = 2,	
\gamma_1= \beta_2=0,
\delta_1= -\delta_2 = 4\,a\omega,
\eta_1 (\omega)= \eta_2 (-\omega)=1-E-2\,a\omega-{a}^{2}{\omega}^{2}$

{\bf The equation governing the spectrum of the TRE is}:
\begin{align}
&R(r)\!=\!C_1R_1(r)+C_2R_2(r), \text{for} \label{R2}\\
&R_1(r)=e^{\frac{\alpha\,z}{2}}(r\!-\!r_+)^{\frac{\beta\!+\!1}{2}}(r\!-\!r_-)^{\frac{\gamma\!+\!1}{2}}\text{HeunC}(\alpha,\beta,\gamma,\delta,\eta,z)\!\notag\\
&R_2(r)=e^{\frac{\alpha\,z}{2}}(r\!-\!r_+)^{\frac{\!-\!\beta\!+\!1}{2}}(r\!-\!r_-)^{\frac{\gamma\!+\!1}{2}}\text{HeunC}(\alpha,\!-\!\beta,\gamma,\delta,\eta,z),\notag
\end{align}
\noindent where $z=-\frac{r-r_+}{r_+-r_-}$ and the parameters are:

$\alpha =-2\,i \left( {\it r_{_+}}-{\it r_{_-}}
 \right) \omega$,  $\beta =-{\frac {2\,i(\omega\,({a}^{2}+{{\it
r_{_+}}}^{2})+am)}{{\it r_{_+}}-{\it r_{_-}}}}-1$, $\gamma ={\frac {2\,i(\omega\,({a}^{2}+{{\it
r_{_-}}}^{2})+am)}{{\it r_{_+}}-{\it r_{_-}}}}-1$,\\
$\delta =-2i\!\left({\it r_{_+}}-{\it r_{_-}}
\right)\!\omega\!\left(1-i
 \left( {\it r_{_-}}+{\it r_{_+}} \right) \omega \right)$,\\
$\eta =\!\frac{1}{2}\frac{1}{{ \left({\it r_{_+}}\!-\!{\it r_{_-}}
\right) ^{2}}}%\times\\
\Big[ 4{\omega}^{2}{{\it r_{_+}}}^{4}\!+ 4\left(i \omega
\!-\!2{\omega}^{2}{\it r_{_-}}\right) {{\it r_{_+}}}^{3}\!+\! \left(
1\!-\!4a\omega\,m\!-\!2{\omega}^{2}{
a}^{2}\!-\!2E \right) \times \\ \left( {{\it r_{_+}}}^{2}\!+\!{{\it r_{_-}}}^{2}
\right) \!+\! %\\
 4\left(i\omega\,{\it r_{_-}} \!-\!2i\omega\,{\it r_{_+}}\!+\!E\!-\!{\omega}^
{2}{a}^{2}\!-\!\frac{1}{2} \right) {\it r_{_-}}\,{\it r_{_+}}\!-4{a}^{2} \left(
m\!+\!\omega\,a
\right) ^{2} \Big].$

Here the $\text{HeunC}(\alpha,\beta,\gamma,\delta,\eta,z)$ is the confluent Heun function -- the unique local Frobenius solutions of the second-order linear ordinary differential equation of the Fuchsian type \cite{heun3_,heun1_},\\ \cite{heun2_} with 2 regular singularities ($z=0,1$) and one irregular ($z=\infty$), normalized to 1 at $z=0$ (\cite{Fiziev1,Fiziev3,spectra,arxiv3},\\ \cite{arxiv1}). The singularities of the TAE are the poles of the sphere (regular) and infinity (irregular), those of the TRE are the two horizons: $r=r_{\pm}$ (regular) and $r=\infty$ (irregular).

\section{Boundary conditions}
In order to obtain the spectrum, one needs to impose specific boundary conditions.

For the angular equation, one requires angular regularity. The Wronskian of two solutions regular around one of the poles, $S_1(\theta)$ and $S_2(\theta)$, should become equal to zero, i.e.  $W[S_1(\theta),S_2(\theta)]=0$, i.e.

\begin{align}
 W[S_1,S_2]=\frac{\text{HeunC}'(\alpha_1,\beta_1,\gamma_1,\delta_1,\eta_1,\left( \cos \left( \pi/6  \right)  \right) ^{2})}{\text{HeunC}(\alpha_1,\beta_1,\gamma_1,\delta_1,\eta_1,\left( \cos \left( \pi/6  \right)  \right) ^{2})}+\notag\\
\frac{\text{HeunC}'(\alpha_2,\beta_2,\gamma_2,\delta_2,\eta_2,\left( \sin \left( \pi/6  \right)  \right) ^{2})}{\text{HeunC}(\alpha_2,\beta_2,\gamma_2,\delta_2,\eta_2,\left( \sin \left( \pi/6  \right)  \right) ^{2})}+ p=0
\label{Wr1}
\end{align}
\noindent where the derivatives are with respect to $z$, and we use $\theta=\!\pi\!/\!3$ (the QNM should not depend on the value of the parameter $\theta$ in the spectral conditions). The parameter ''p`` can be found in \cite{arxiv2014}.

For the radial equation, one can impose the black hole boundary conditions (BHBC): waves going simultaneously into the event horizon ($r_+$) and into infinity; or the quasibound boundary conditions (QBBC): waves going simultaneously out of the event horizon ($r_+$) and out of infinity \cite{spectra, arxiv2014}. Those two conditions lead to:

-- BHBC: For $m=0$, the only valid solution is $R_2$, while for $m\neq0$, the solution $R_2$ is valid for frequencies for which $\Re(\omega) \not\in (-\frac{ma}{2Mr_+},0)$. Also, it requires that: $\sin(\arg(\omega)\!+\!\arg(r))\!<0$ (the direction of steepest descent).

-- QBBC: For $m=0$, the only valid solution is $R_1$, while for $m\neq0$, the solution $R_1$ is valid for frequencies for which $\Re(\omega) \not\in (-\frac{ma}{2Mr_+},0)$. Also, it requires that: $\sin(\arg(\omega)\!+\!\arg(r))\!>0$.

The total-transmission modes (TTM) do not appear in the EM case, in contrast to the gravitational case.

Here, we will use the $\epsilon$-method \cite{arxiv2014} in its most direct form $r=|r|e^{i\arg(r)}$ to vary $\arg(r)$, and afterwards we will check whether the BHBC or the QBBC are satisfied.

\section{Numerical methods}
In order to obtain the spectrum $\omega_{n,m}(a)$, we will set the mass to  $M=1/2$ and the radial infinity to $|r|=110$ (where not stated otherwise) and we will solve Eqs.\eqref{Wr1} and \eqref{R2}. The system of two equations featuring the confluent Heun functions will be solved by the two-dimensional M\"uller algorithm, developed by the team \cite{arxiv3, arxiv1}. The algorithm is realized in \textsc{maple} code and the numbers presented below are obtained using \textsc{maple} 18 and 2015. The software floating point number (the parameter ''Digits``) is set to 64 (unless stated otherwise), the precision of the algorithm -- to 25 digits.

Note that the numerical error due to the integration is hidden in the evaluation of the confluent Heun function which depends on the Maple parameter ''Digits``. With current algorithms, one is not able to set the final number of stable digits in the HeunC evaluation, but one can request the integration to be performed with certain precision, controlled by ''Digits``.
In general, with our choice of the parameters of the root-finding algorithms, we can expect at least 15 digits of precision of the final result for the QNM/QBM, a number which can be increased by increasing further ''Digits``, but increasing also the computational cost of the calculations.

\section{The stability of the numerical spectra}
The numerical results of the QNM and QBM spectra are presented in \\
\cite{arxiv2014}. Here we will focus only on the tests of the numerical stability of those modes. We will consider two types of deviations: deviation in the absolute value of the radial variable and deviations in the argument of radial variable. The QNM/QBM spectra should not depend on the radial variable at all, because $\omega$ is a separation constant of the Teukolsky Master Equation. If the spectra depend on $r$, then either the actual numerical infinity is too low and we need to go further away from the horizon (i.e. increasing $|r|$), or the spectrum is not QNM/QBM, but a numerical artifact. One should keep in mind that the QNM/QBM are expected to be stable on $\arg(r)$ only in a limited region around the direction of steepest descent \cite{arxiv2014}, which in principle can be a curve and not a straight line, as in the first approximation.

As discussed in \cite{arxiv2014}, we obtained 3 types of spectra: the QNM, the QBM and a spurious spectrum. The last one is not independent on the value of the radial variable, as we demonstrate below. Thus,  it cannot be considered a physical solution of the spectral problem, even though its $\{\omega_{ln}, E_{ln}\}$ are formal roots of the spectral system.

\subsection{The QNM/QBM spectrum}

To study numerically the stability of the QNM/QBM spectra, we will work only with the $a=0$ case (no rotation) because it is computationally simpler. The results are readily extended to the case with rotation ($a>0$).

For the non-rotating case, the spectra are found directly from the radial equation -- $R_2 = 0$ ($R_1 = 0$) for the QNM (QBM) -- since in this case the angular solutions are known: $E=l (l+1)$, $l=1,2..$. This means that we can work with the one-dimensional M\"uller algorithm, used very successfully in the case of the jet-spectra of the black hole and naked singularities \cite{spectra}. The main difficulty encountered in our previous works was the limited precision of the algorithms evaluating the confluent Heun functions. In Maple version 18 and higher, those algorithms have been redesigned to allow arbitrary precision for HeunC. This facilitated significantly the current stability study and because of it, now it is possible to obtain the QNM and the QBM with arbitrary digits of significance.

The TRE, just as the TAE, has 16 classes of solutions, valid in different sectors of the complex plane. While for the QNM the solution $R_2(r)$ covers the needed sectors of the complex plane, for the QBM the above solution $R_1(r)$ requires a significant increase in the parameter Digits compared to $R_2(r)$ to obtain the same precision. For this reason, we used one of the other, better convergent, solutions of the TRE which are linearly independent of $R_2(r)$ and have the same roots. The new solution differs from $R_1(r)$ only because in it $\alpha\to -\alpha$.
Similarly, one may find another solution which produces the QNM and it differs from $R_2(r)$ by change $\gamma\to-\gamma$. Over all, there are 4 solutions from which one can obtain the QNM spectra and 4 for the QBM spectra, the only difference being the convergence of the root-finding algorithm.

\begin{figure}[!ht]
\vspace{-0cm}
\centering
\subfigure[\, m=0, l=1, 2]{\includegraphics[width=150px,height=130px]{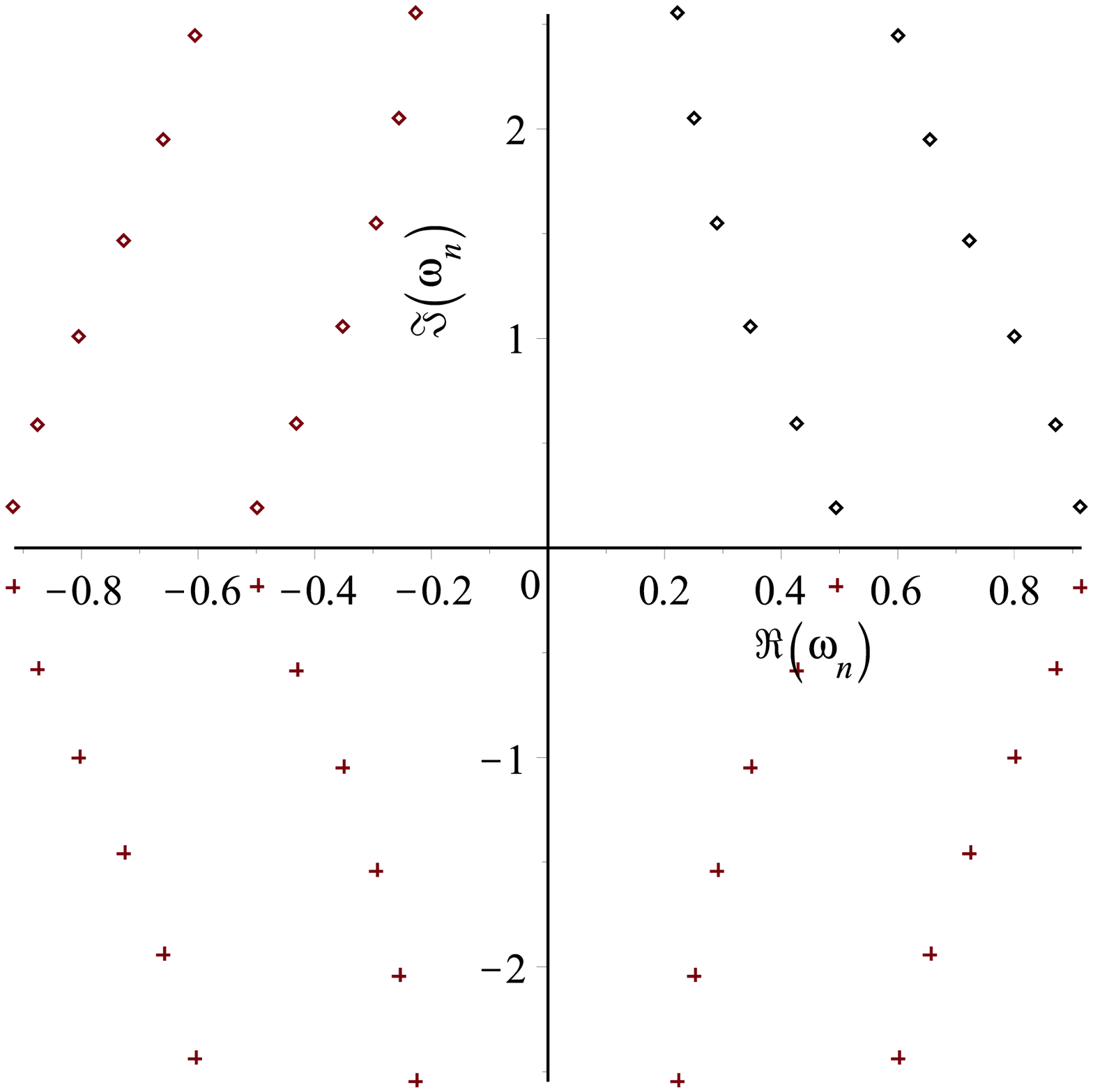}}
\subfigure[\,  m=0, l=1, 2]{\includegraphics[width=150px,height=130px]{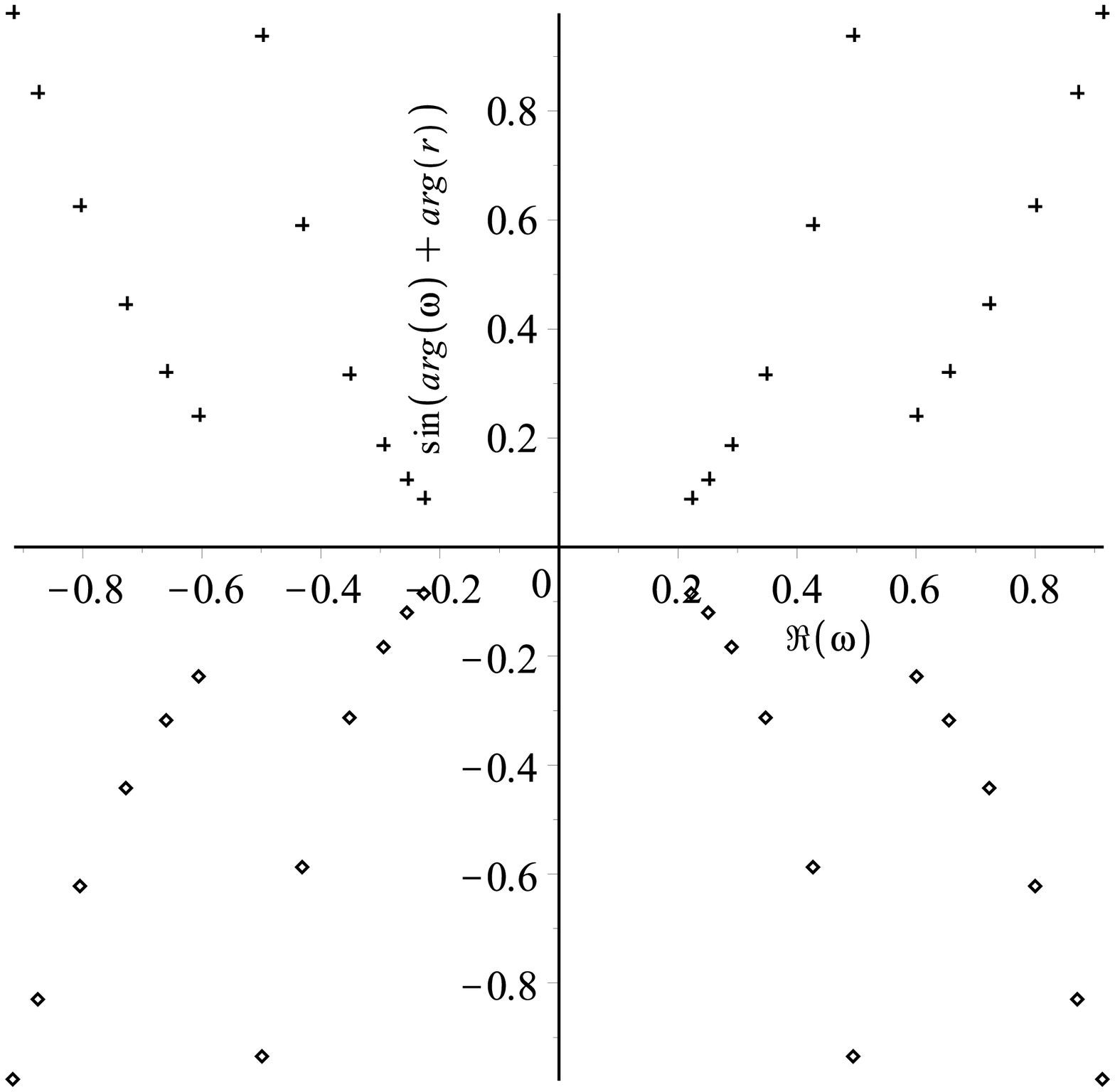}}
\caption{(a) QNMs (diamonds) and QBMs (crosses) for $a=0$,$m=0, l=1, 2, n=0..5$. (b) the boundary condition $\sin(\arg(\omega)+\arg(r))$ for both the  types of spectra. }
\label{Fig1}
\end{figure}

As mentioned above the QNM/QBM are a discrete set of frequencies numbered $n=0,1, ...$ in the direction of increasing absolute value of their imaginary part, thus denoted here as $\omega_n$. One can see the QNM/QBM frequencies with $n=0..5$ on Fig. \ref{Fig1} a). On Fig. \ref{Fig1} b) we have plotted the boundary conditions corresponding to the 2 types of spectra.

\begin{figure}[!ht]
\vspace{-0cm}
\centering
\subfigure[\, m=0, l=1]{\includegraphics[width=120px,height=120px]{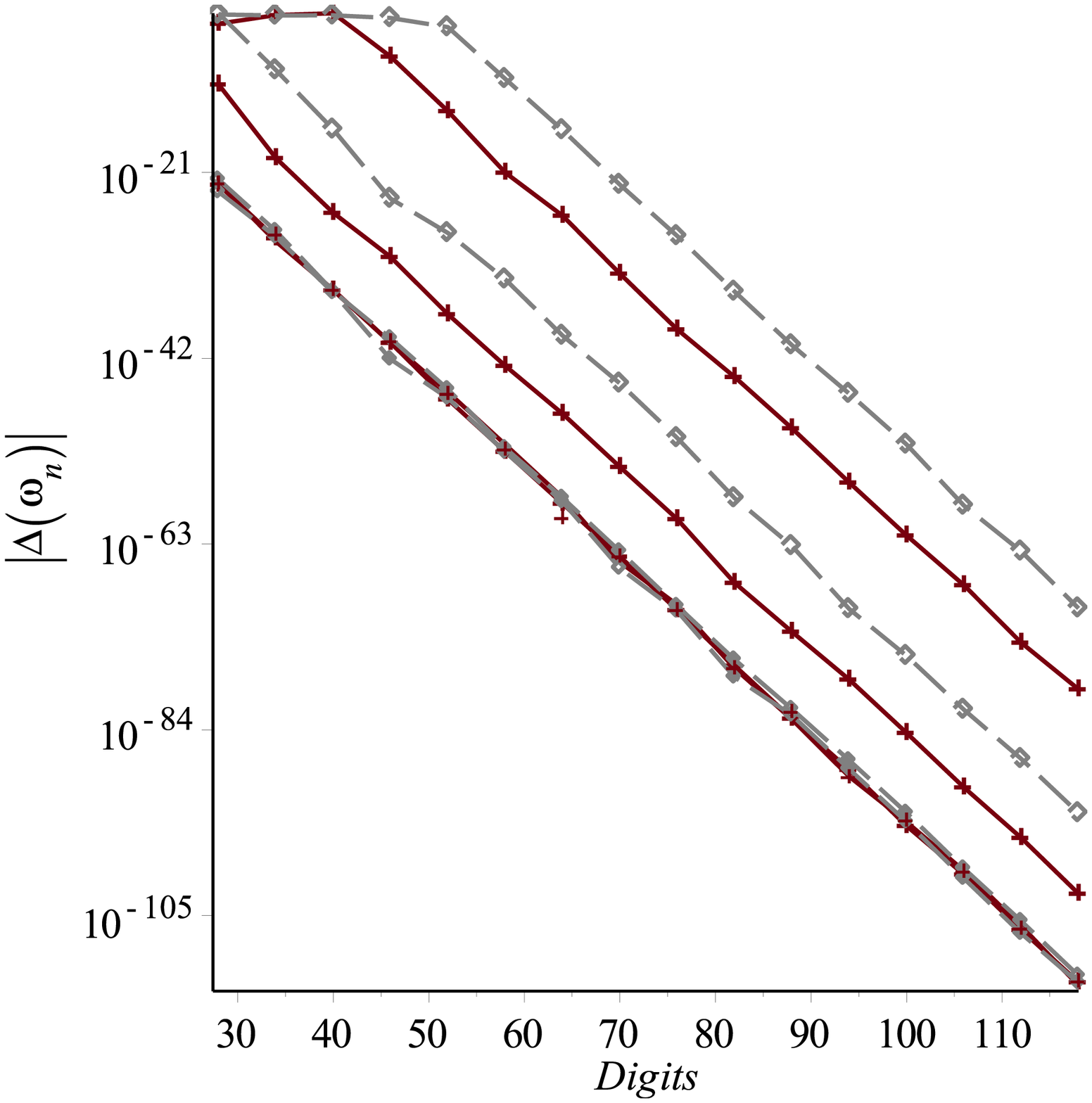}}
\subfigure[\, m=0, l=1]{\includegraphics[width=120px,height=120px]{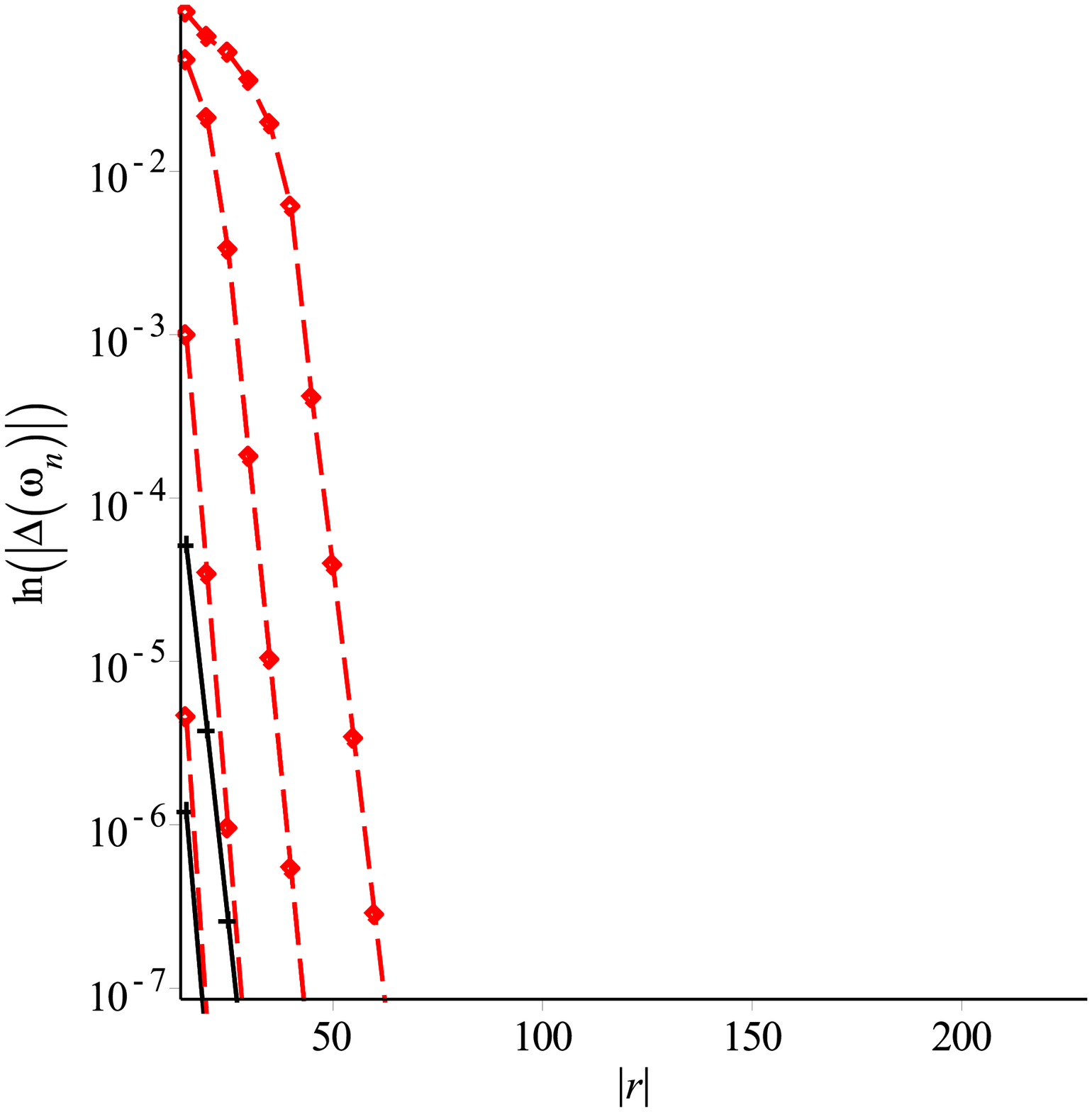}\llap{\raisebox{1.5cm}{\includegraphics[height=2.5cm]{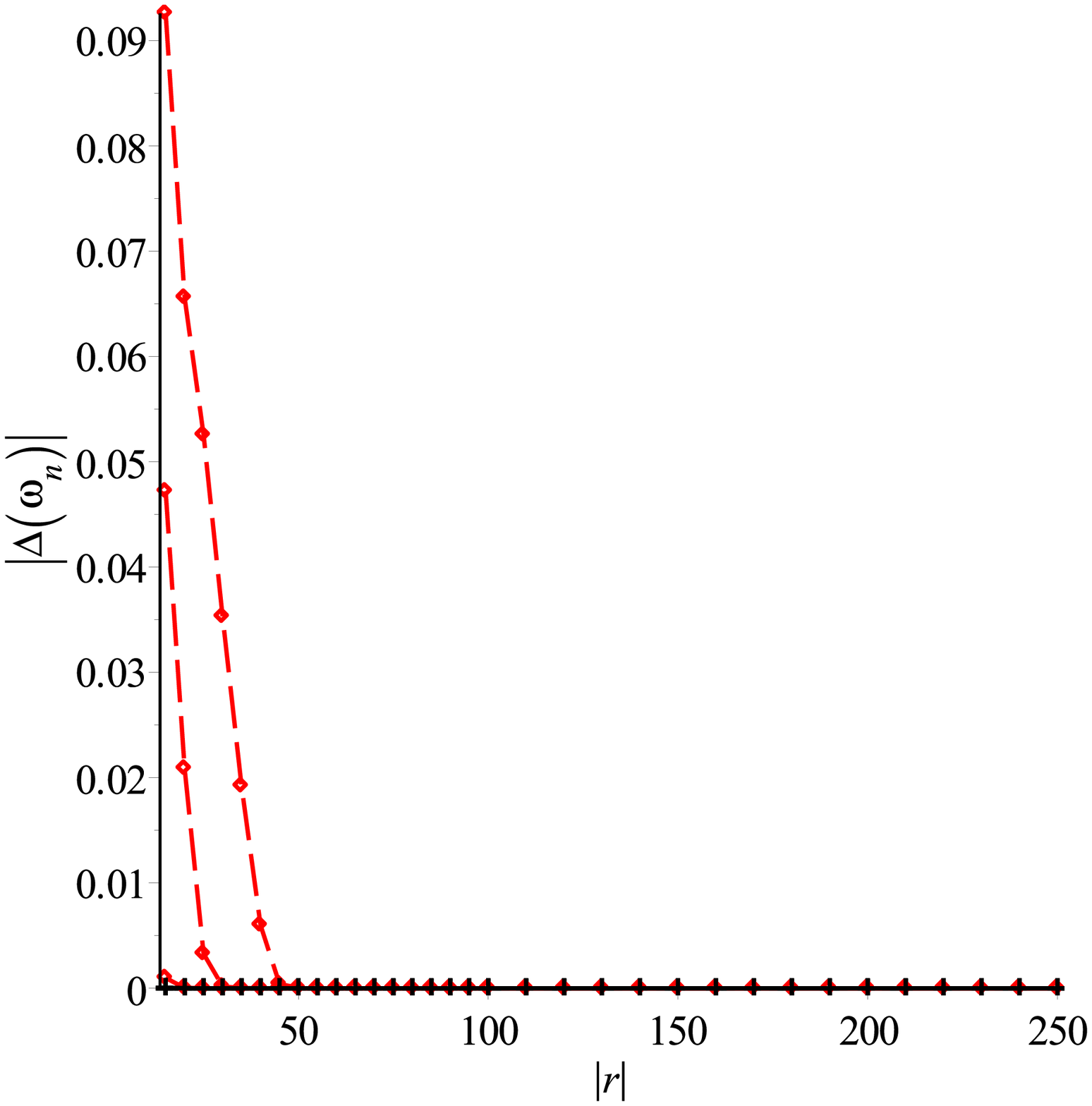}}}}
\subfigure[\, m=0, l=1]{\includegraphics[width=120px,height=120px]{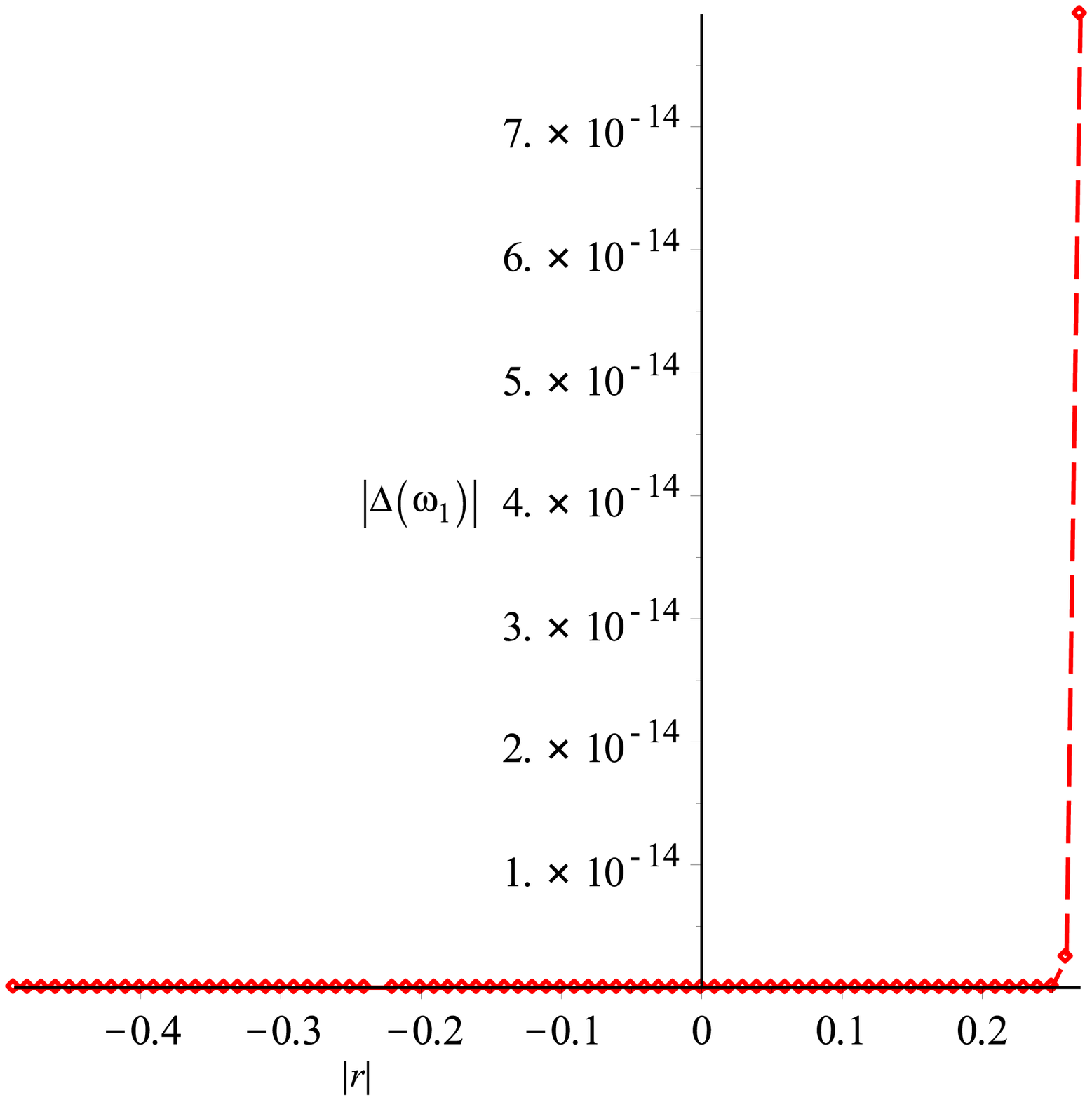}}
\caption{(a) The dependence of the QNMs (diamonds) and QBMs (crosses) on the parameter Digits for $a=0$, for $m=0, l=1, n=0..3$ (log-scale). (b) The dependence of the QNMs (dash line) and QBMs (solid line) on the radial infinity $|r|$ for $a=0$, for $m=0, l=1, n=0..3$ for $|r|=10..230$ (c) The dependence of one frequency ($n=0$) on the parameter $\epsilon$ for $\epsilon=-0.5..0.27$ . }
\label{Fig2}
\end{figure}

\begin{figure}[!ht]
\vspace{-0cm}
\centering
\subfigure[\, m=0, l=1]{\includegraphics[width=120px,height=120px]{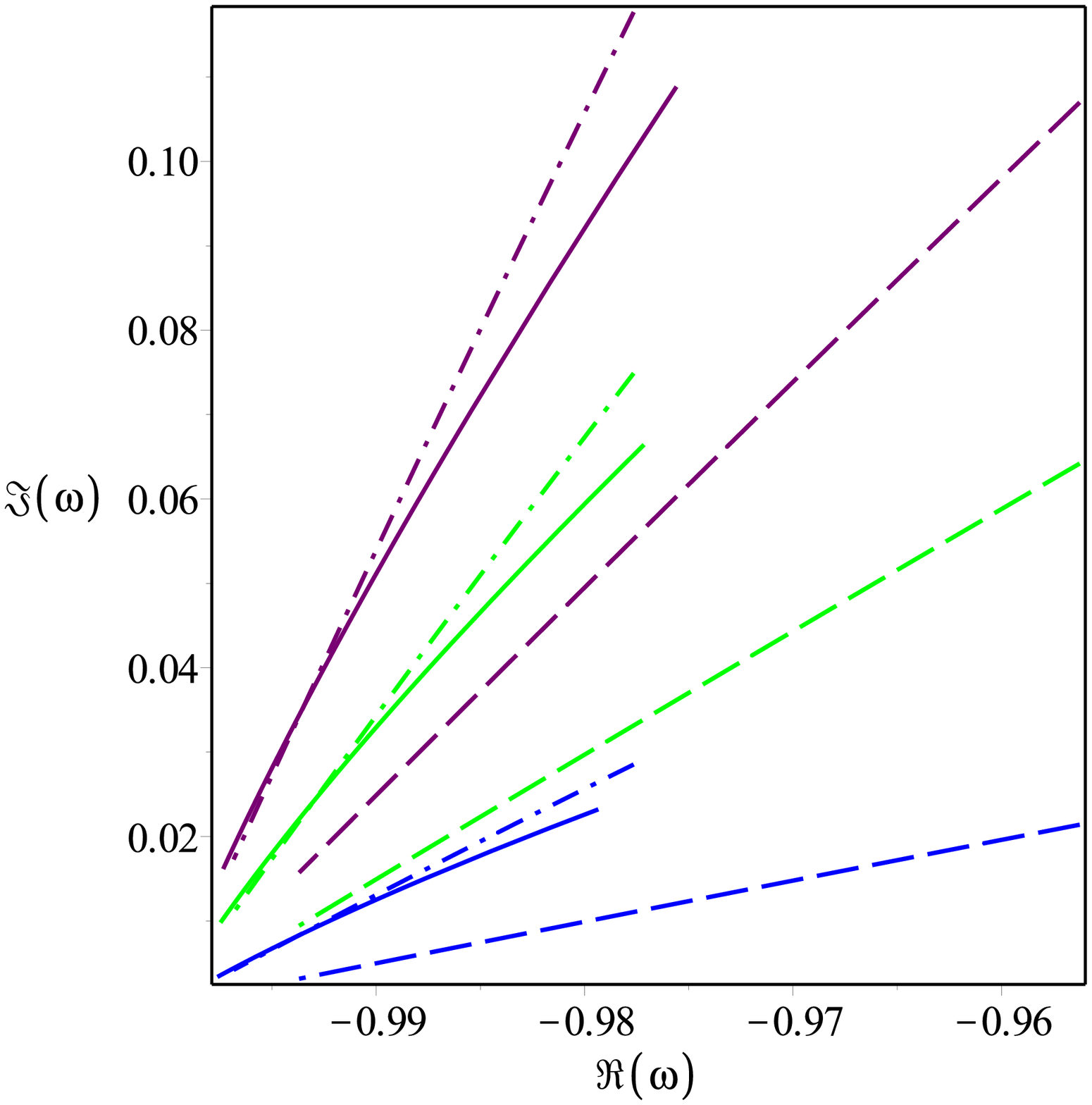}}%Fig1_b
\subfigure[\, m=0, l=1]{\includegraphics[width=120px,height=120px]{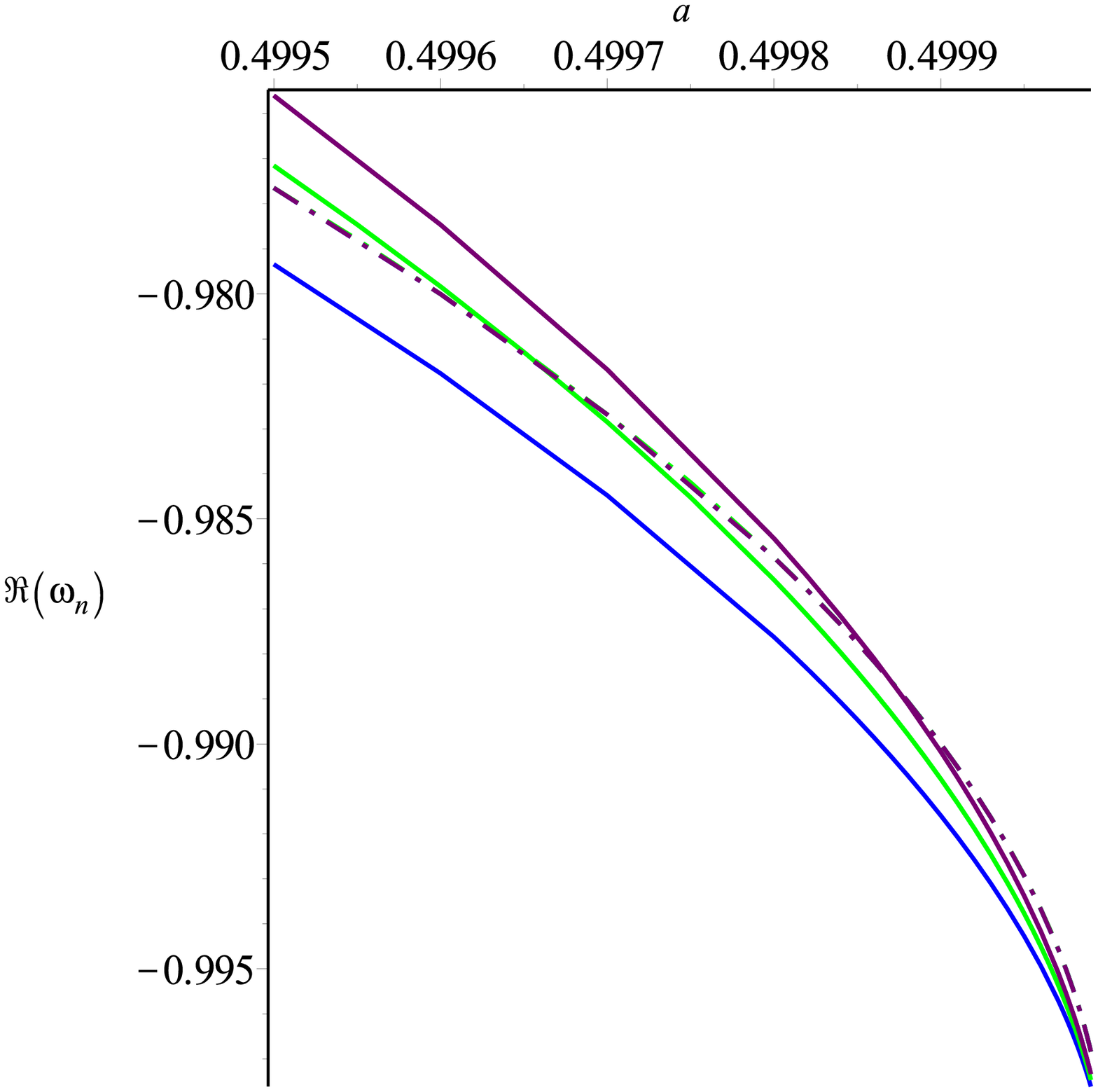}}
\subfigure[\, m=0, l=1]{\includegraphics[width=120px,height=120px]{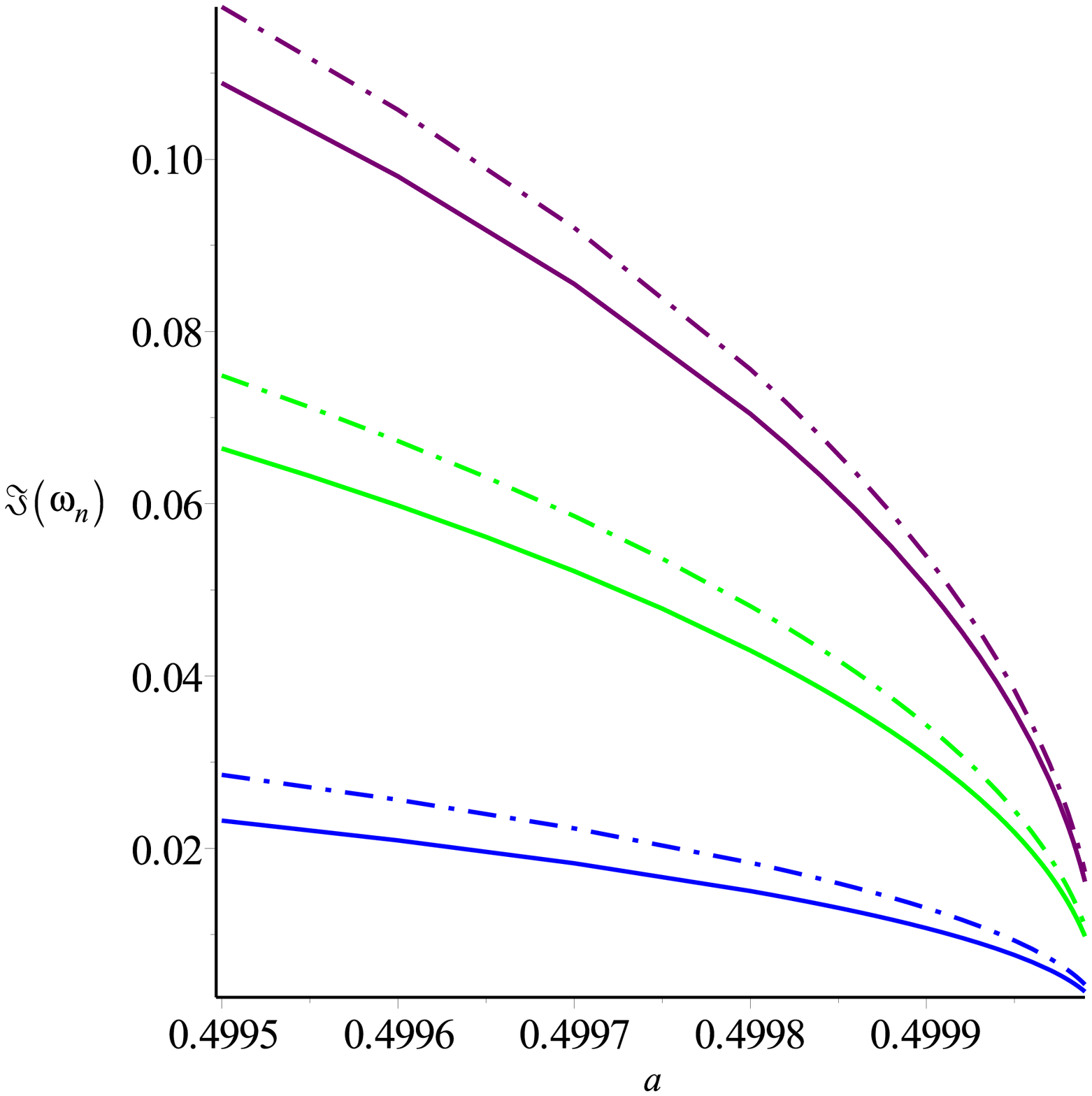}}
\caption{(a) Complex plot of the approximation of $\omega^1(a)$ (dashed line), $\omega^2(a)$ (dot-dashed line), and our numerical results $\omega_n$ for n=0..2, m=1, l=1 b) and c) $\Re(\omega)(a)$ and $\Im(\omega)(a)$  for our results and $\omega^2(a)$ (the legend is the same). }
\label{Fig3}
\end{figure}

The study of the stability in this case can be seen on Fig. \ref{Fig2}-\ref{Fig4}. On them, we focused on the following cases:
\begin{enumerate}
 \item Stability of the modes with respect to the parameter Digits in Maple

  The results are on Fig. \ref{Fig2} a) where on the x-axis we have plotted the ''Digits`` in use and on the y-axis the real part of the difference between two consecutive frequencies evaluated for those Digits: $\Re(\omega_i-\omega_{i-1})(Digits)$ for 4 different modes $n$. Clearly the difference quickly falls to a very small number and for Digits>60 for $n=4$, one has $\Re(\omega_i-\omega_{i-1})<10^{-23}$. The general tendency is that one needs higher Digits for higher modes.
 \item Stability with respect to the absolute value of the radial variable

 On Fig. \ref{Fig2} b) we have plotted $|\Delta(\omega[n])|(|r|)$ in log-scale on the y-axis, where $\Delta(\omega[n])=\omega[n](r)-\omega[n](r')$. From the plot it is clear that the frequencies depend on $|r|$ only near the horizon, with higher modes depending on $r$ stronger. For the modes on the plot ($n=0..3$), $|r|>90$ corresponds to the actual radial infinity after which the modes stabilize and remain the same with further increase of $|r|$ (with the difference between two consecutive evaluations  $<10^{-14}$).
 \item Stability with respect to the argument of the radial variable

 The QBM and the QNM spectra are stable to the precision required by the root-finding algorithm in a very large interval of epsilon. If we consider $r=|r|e^{\pi(\frac{3}{2}+\epsilon)}$, we find that the QNM frequencies with positive real parts appear for $\epsilon<-0.5$ (i.e. $\arg(r)>\pi$) and remain stable in an interval on $\epsilon$ depending on $n$ -- for $n=0$, $\Delta \omega_{QNM}^{\pm} = (\mp0.5,\pm0.29]$ and for $\Delta \omega_{QBM}^{\mp}= (\mp0.5,\pm0.34]$, for $n=1$, $\Delta \omega_{QNM}^{\pm} = (\mp0.5,\pm0.12]$ and for $\Delta \omega_{QBM}^{\mp}= (\mp0.5,\pm0.17]$, for $n=2$, $\Delta \omega_{QNM}^{\pm} = (\mp0.5,\pm0]$ and for $\Delta \omega_{QBM}^{\mp}= (\mp0.5,0]$. The symmetry with respect to the imaginary axis in the dependence on $\epsilon$ is clear -- we see that for $\epsilon<-0.5$ the frequencies with the positive real parts disappear and on their place, the root-finding algorithm finds the QNM frequencies with the negative real parts. For the QBM frequencies we observe the inverse behavior -- those with the negative real part appear for $\epsilon>-0.5$. This symmetry is due to the fact that the $\omega^{\pm}$ frequencies of each type are obtained with the same differential equation in a different region of the complex plane.

 Additionally, one can see that the interval on $\epsilon$ moves to the left with the increase of $n$, which, combined with the need of increasing of the Digits makes the detailed study of the $\epsilon$-dependence for the modes with $n>4$ computationally expensive. On Fig. \ref{Fig2} c)  the dependence $\omega_n(\epsilon)$ for $n=0$ is plotted, but it is representative for the behavior of all modes. Generally, what we observe is a plateau-like region of the $\epsilon$-interval where the frequencies remain stable with more than 20 digits, followed by a region where the frequencies start changing smoothly, but in a very steep manner.
 \end{enumerate}

All the numerical data point to the stability of the so-obtained frequencies in certain intervals on $\epsilon$ and $|r|$, just as expected by the theory.

{\bf Near-extremal regime ($a\to M$)}

Let us define the angular velocity of the KBH: $\Omega=\frac{a}{r_+^2-a^2}$, with $r_{\pm}=M\pm\sqrt{M^2-a^2}$ being the inner and the outer BH horizon; the area of the KBH horizon: $A=4\pi(r_++a^2)$ and the Hawking temperature: $T_{BH}=\frac{r_+-r_-}{A}$.  Hod in \cite{EBH} derived two analytical formulas for the frequency $\omega$ in the near-extremal regime in terms of $\Omega$ and $T_{BH}$, i.e. $\omega^1=m\Omega-i2\pi T_{BH} (n+1/2)$ and $\omega^2=m\Omega-i2\pi T_{BH} (n+1/2+i \delta)$, where $\delta$ is a complex number. As part of our test of the validity of our numerical results, we tested how well this theoretical formula approximates the so-obtained frequencies. The results can be seen on Fig. \ref{Fig3}.

Clearly, $\omega^2$ fits the numerical results better. On those plots we have used $\delta = -1/2+i/6$, although for the case $m=2$ the best fit is for $\delta = -1/2$ . The deviation from the theoretical formula is less than $5\%$ for $a \to M$, which is surprising considering the fact that the confluent Heun function in this regime is expected to become numerically unstable. The fact that $\omega^1$ does not approximate the frequencies better, even though $m=l$, may be due to the low $n$ we are using, since the Hod's original approximation was for infinitely damped frequencies (i.e. $n$ very high).

{\bf QNMs in physical units}

Finally, we would like to calculate the QNM frequencies in physical units for two observationally well-studied objects. From
\cite{BHrot} we know that: LMC X-1 has $a_*^1=0.92$,  $M^1=10.91\Msol$ and Cygnus X-1 -- $a_*^2=0.983$,  $M^2=14.8\Msol$. Using the formula provided in the Appendix, we find the physical frequencies of the first two QNMs to be:

\noindent$\omega_{n=0,m=1}^{1,2}=1.178kHz, 0.976 kHz, \tau_{damp}^{1,2}=0.86ms, 1.80 ms$,
\\ $\omega_{n=1,m=1}^{1,2}=1.160kHz, 0.968 kHz, \tau_{damp}^{1,2}= 0.29ms, 0.61 ms$.

Clearly, frequencies of less than a kHz damping in a manner of milliseconds, are inaccessible to even the best Earth radio telescopes because of the ionospheric absorption and scattering. A possible lunar or space-based radio telescope could listen below $30$MHz, but even then, hearing the ringing of stellar black holes would be extremely difficult because of their matter-rich surrounding which will absorb the EM spectrum. Such observations can be made only if the QNMs are up-scattered by some unknown mechanism to frequencies observable on the Earth. This is an open question facing theoretical and observational astrophysics.

%Also, this low frequency window from ~30 kHz (just above the local plasma frequency of the interplanetary medium) to ~30 MHz (where high resolution observations from the ground become possible most of the time) spans three orders of magnitude in frequency, wider than the infrared window opened by IRAS and ISO or the ultraviolet window opened by IUE and EUVE. It is the last region of the electromagnetic spectrum which is still largely unexplored.%http://www.nrl.navy.mil/rsd/7210/7213/lfra-from-space

\subsection{The spurious spectrum}
\begin{figure}[!ht]
\vspace{-0cm}
\centering
\subfigure[\,  m=0, l=1, 2]{\includegraphics[width=150px,height=130px]{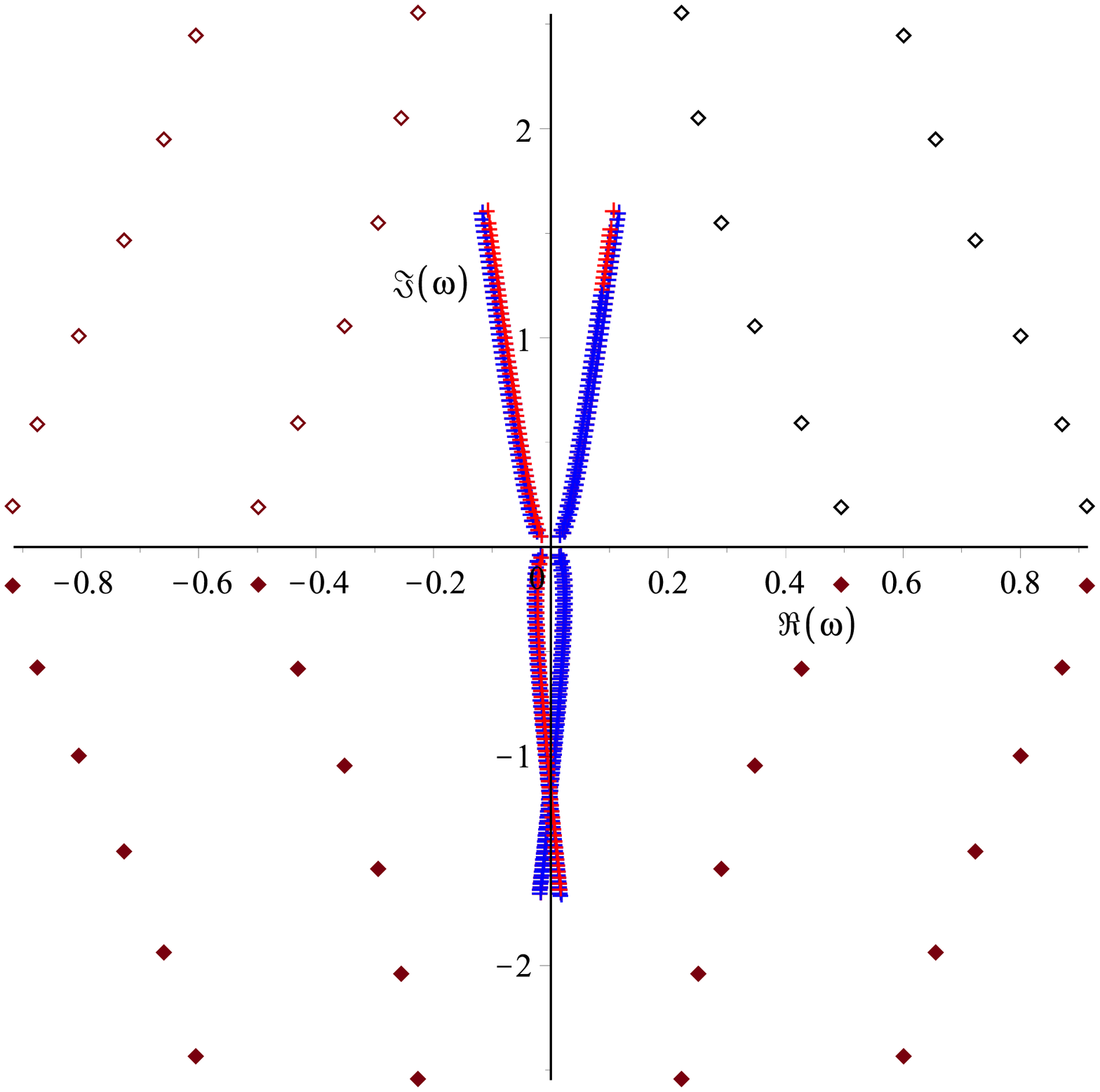}}%Fig1_b
\subfigure[\, m=0, l=1, 2]{\includegraphics[width=150px,height=130px]{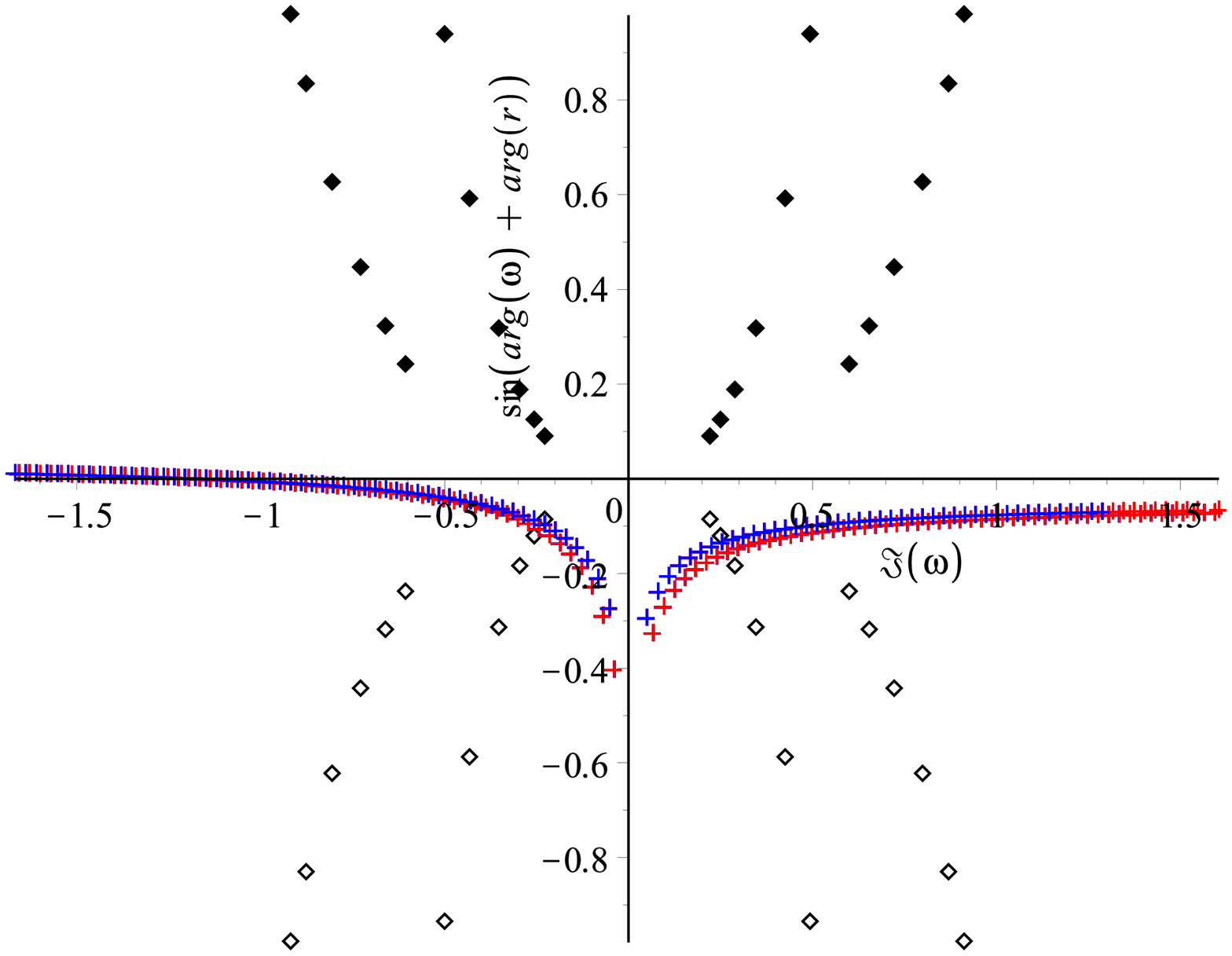}}
\caption{(a) The spurious spectrum (crosses) plotted next to the QNM and QBM spectra (diamonds) for $a=0$, for $m=0, l=1,2$. (b) the corresponding boundary condition $\sin(\arg(\omega)+\arg(r))$ for the two types of spectra -- again the QNMs and QBMs are with diamonds and the spurious spectrum -- with crosses. On the plot, the two spurious spectra for $l=1,2$ are almost coinciding}
\label{Fig4}
\end{figure}

On Fig. \ref{Fig4}, one can see an example of a spurious spectrum corresponding to $l=1,2$. Once again we have a discrete infinite spectrum, although the points on the figure are very closely spaced, and thus the central part looks like a line (see Fig. \ref{Fig4}a). For $l=2$ those frequencies change, like in the QNM/QBM case, however the change is less pronounced. From Fig. \ref{Fig4}b, it is clear that even though the spurious spectrum is very similar to the QNM/QBM, it doesn't follow the same boundary conditions, with some points corresponding to infinity BHBC (QNM) ($\Im(\omega_{sp})>-1.19$) and the rest -- to infinity QBBC (QBM). Because they are roots to both $R_1$ and $R_2$, the boundary condition on $r\to r_+$ is ambiguous --in those points the functions $R_1(r)$ and $R_2(r)$ stop being linearly independent (up to certain numerical precision). 

Those roots have been confirmed with the 3 methods of HeunC evaluation in Maple -- the direct integration, the series expansion and the Taylor expansion. 

\begin{figure}[!ht]
\vspace{-0cm}
\centering
\subfigure[\,]{\includegraphics[width=120px,height=120px]{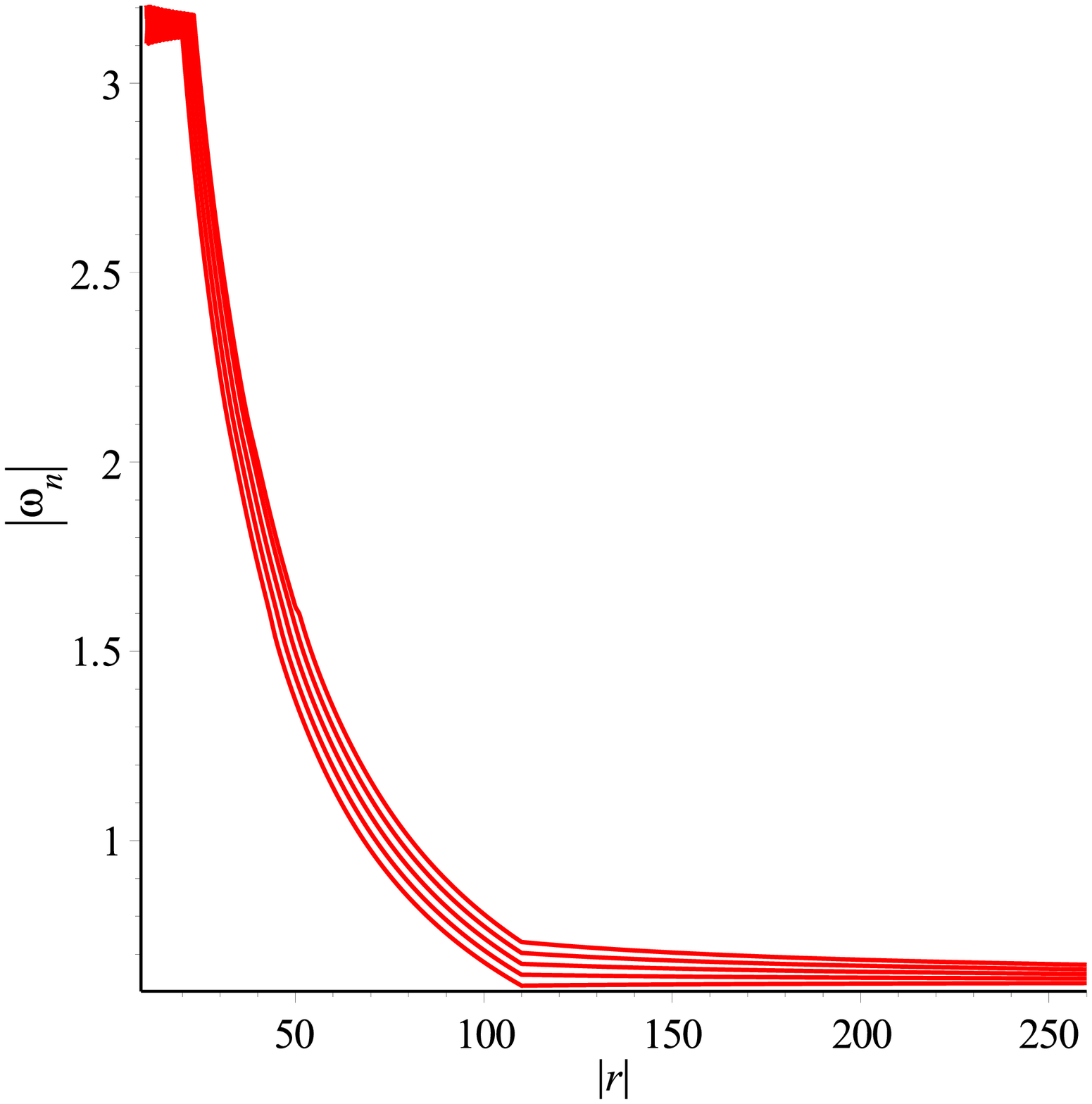}}
\subfigure[\,]{\includegraphics[width=120px,height=120px]{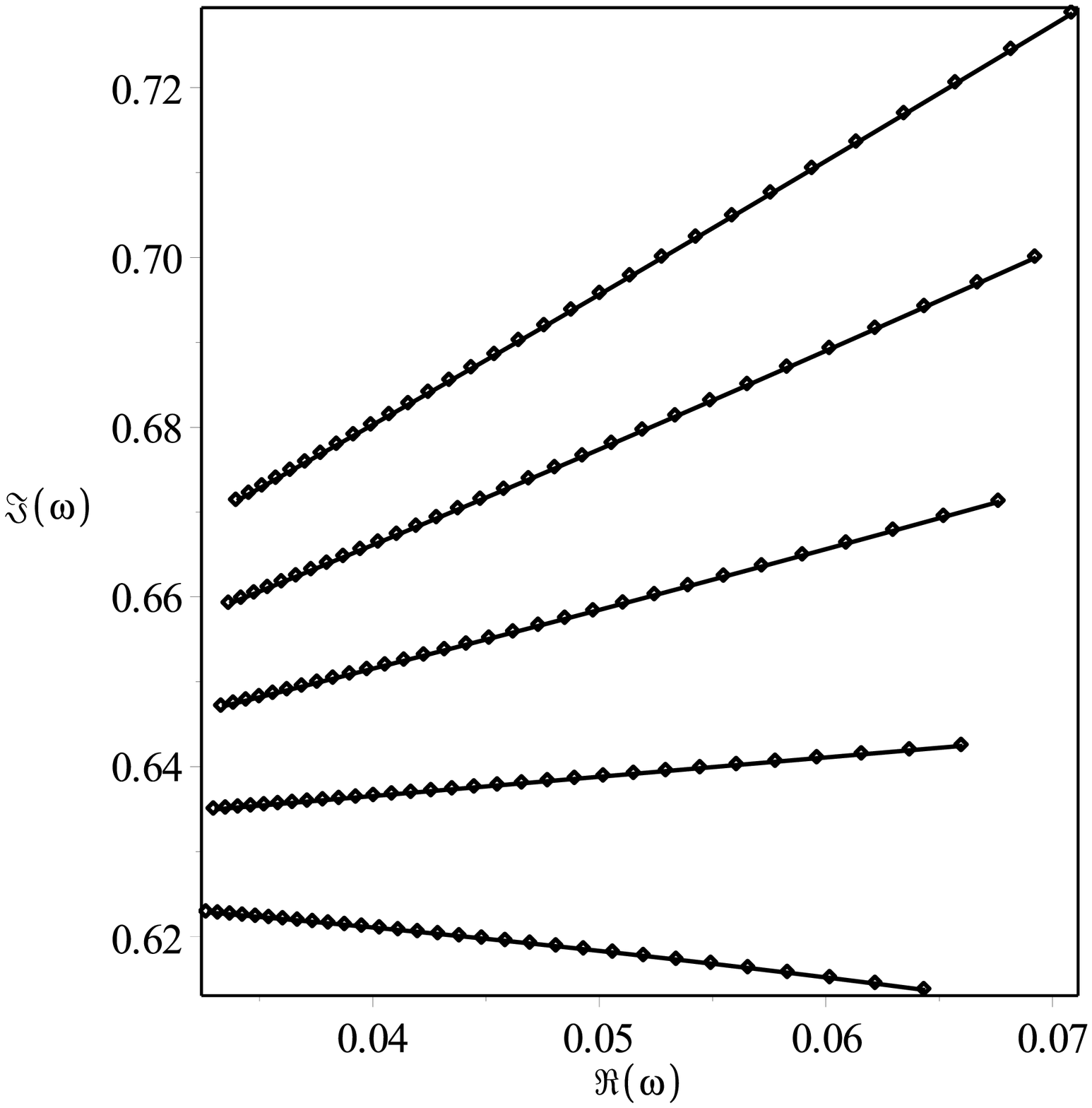}}
\subfigure[\,]{\includegraphics[width=120px,height=120px]{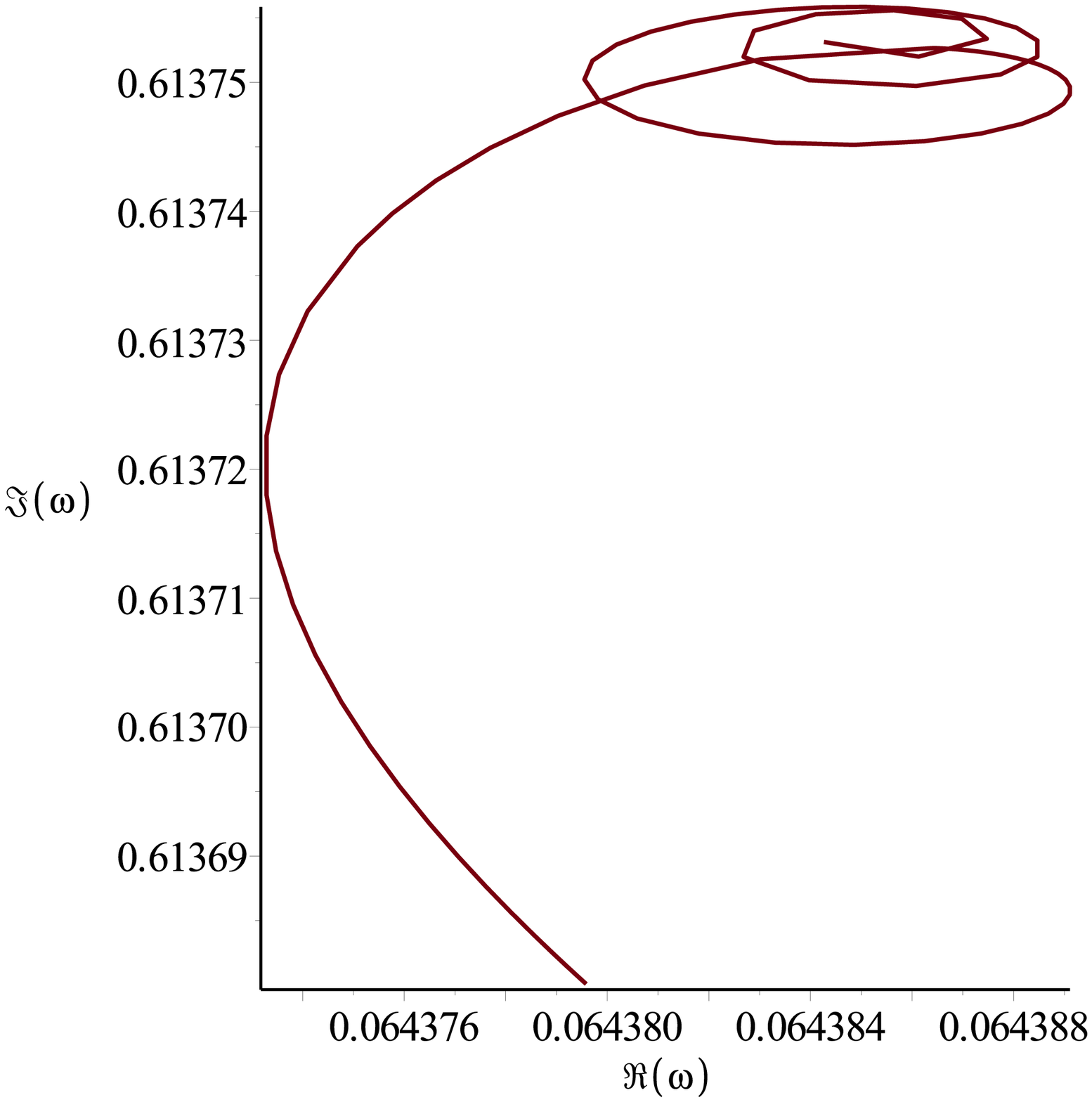}}
\caption{(The dependence of the spurious spectrum on a)  the radial infinity $|r|$ for $a=0$, for $m=0, l=1, n=0..3$ for $|r|=10..250$; (b) a complex plot of 5 consecutive modes $\omega_{m=0,l=1}(|r|)$  for $|r|=110..260$, where $|r|=110$ is to the right   c) a complex plot of $\omega_{m=0,l=1}(a)$ for $a=0..0.495$}
\label{Fig5}
\end{figure}

\begin{figure}[!ht]
\vspace{-0cm}
%\hspace{1.5cm}
\centering
\subfigure[\,]{\includegraphics[width=120px,height=120px]{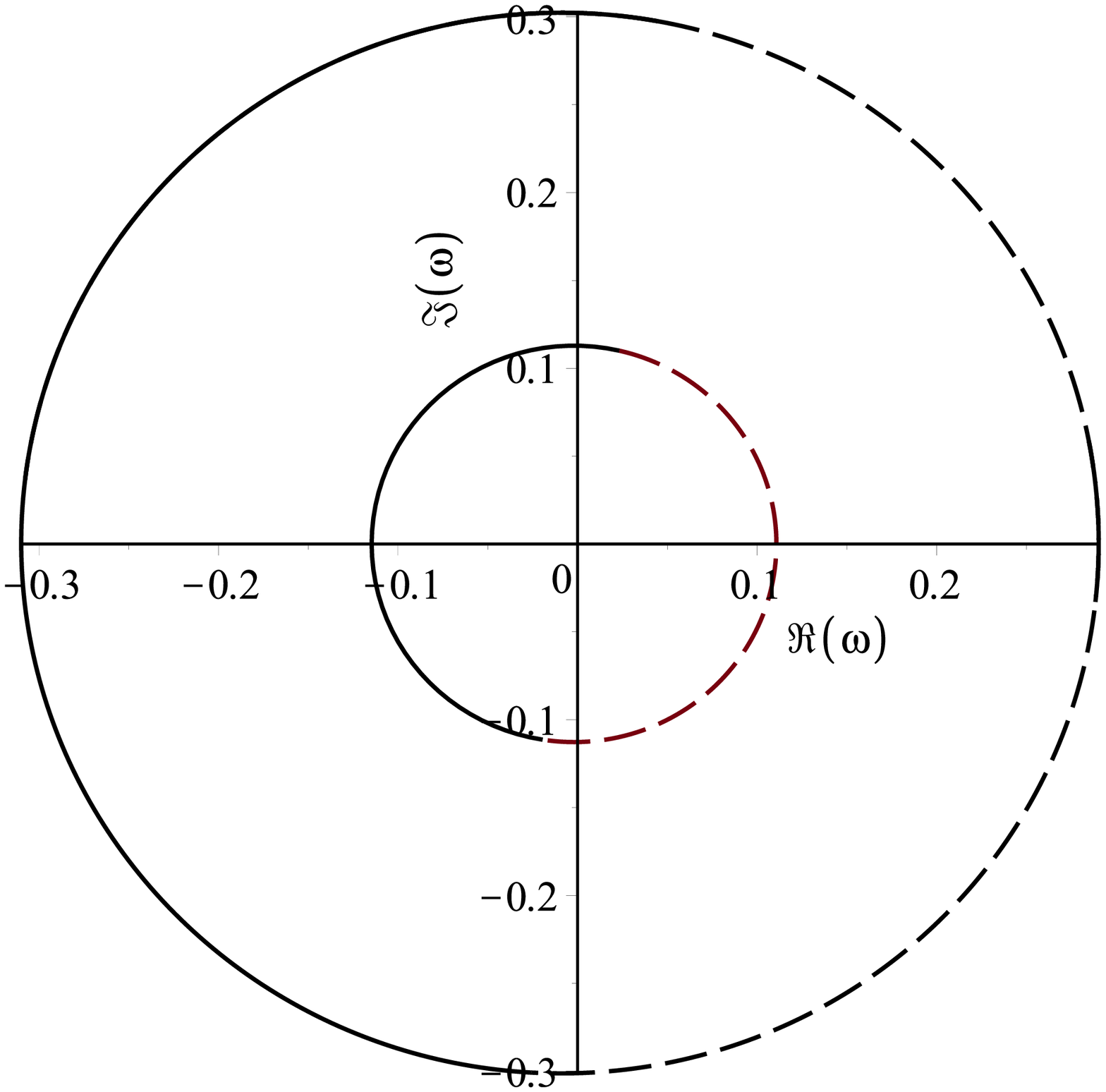}}%Fig1_b
\subfigure[\,]{\includegraphics[width=120px,height=120px]{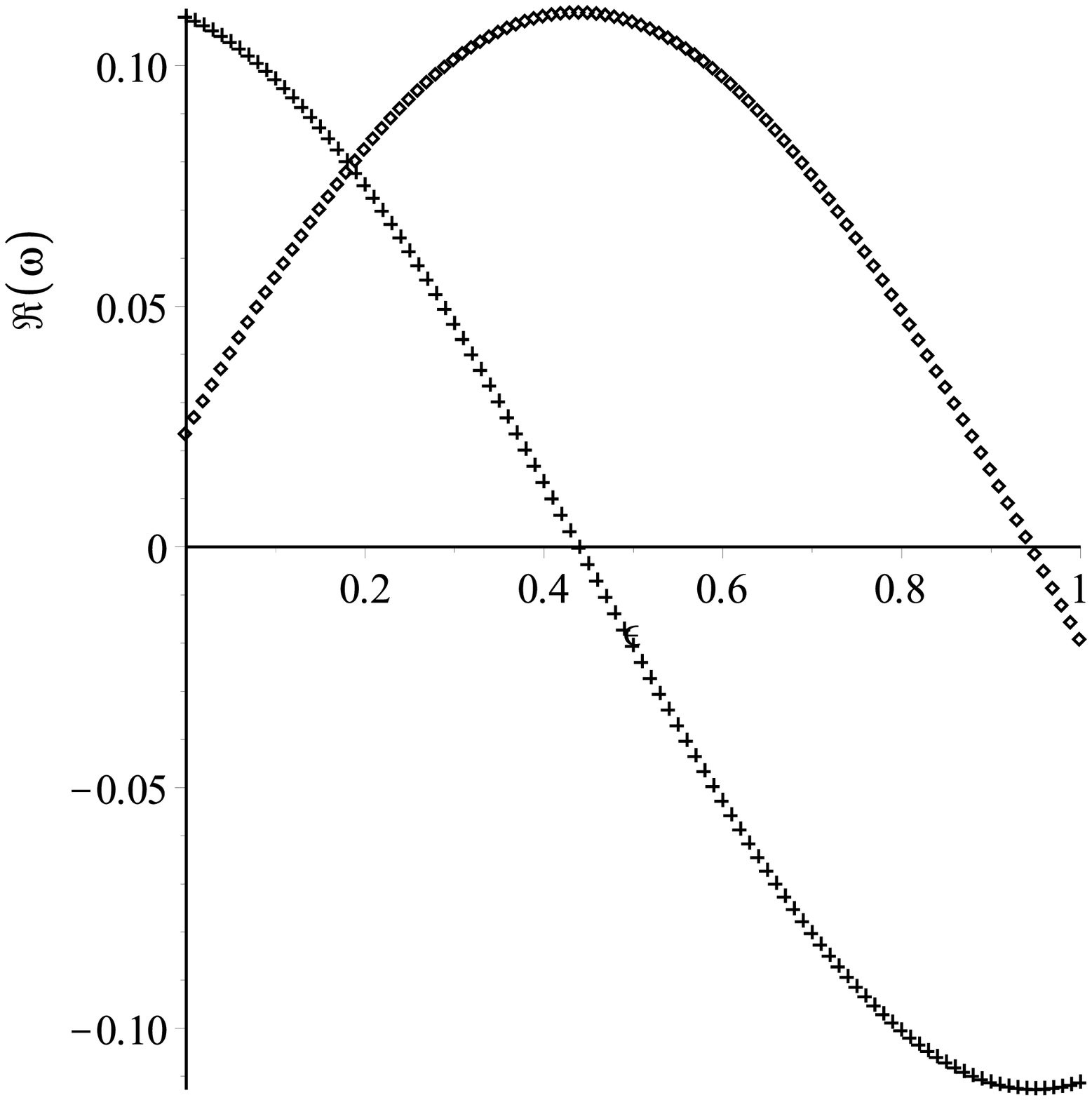}}
\subfigure[\,]{\includegraphics[width=120px,height=120px]{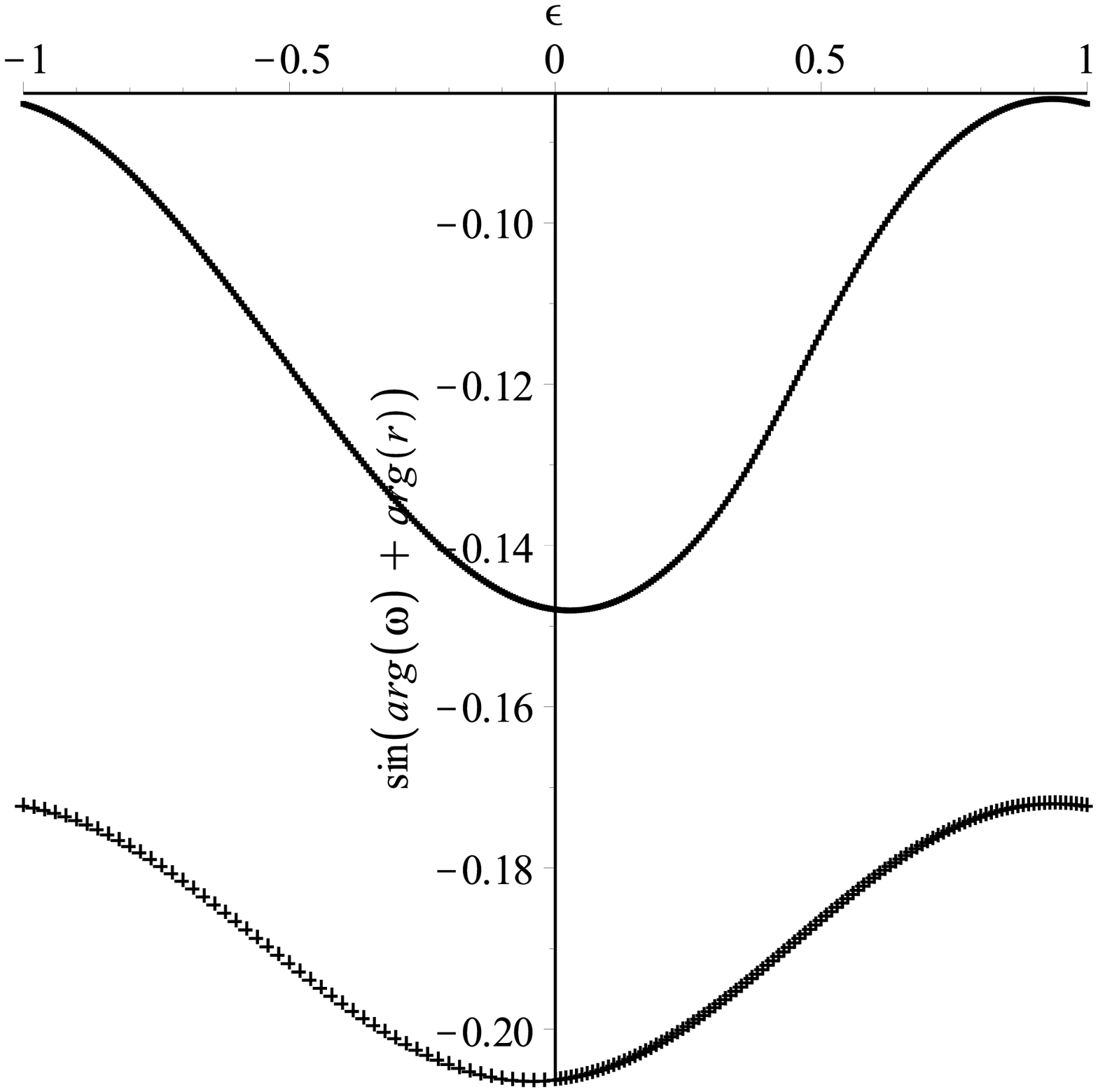}}
\caption{(a)The dependence of the spurious spectrum on the $\epsilon$ parameter: a) a complex plot of the frequencies for $\epsilon=0..\pm1$ -- the positive $\epsilon$ is represented with the dashed line. On the plot the inner circle is the $l=2$ mode, the outer -- the $l=1$ modes. b) the real (diamonds) and imaginary (crosses) parts of the frequencies ($l=2$) as a function of $\epsilon$ c) the boundary conditions for $l=1$ (points) and $l=2$ (crosses)}
\label{Fig6}
\end{figure}

The study on the stability in this case is on Fig. \ref{Fig4}-\ref{Fig6}:
\begin{enumerate}
 \item Stability of the modes with respect to the parameter Digits

 This case repeats our study of the Digits dependence on QNM/QBM, so we did not plot this dependence again. As expected, the modes with a bigger imaginary part require higher value for ''Digits''.
 
 \item Stability with respect to the absolute value of the radial variable

 The results are on Fig. \ref{Fig5} a) and b), where we plotted 5 consecutive spurious modes. On Fig. \ref{Fig5} a) one can see how the absolute value of the frequency corresponding to each mode changes with the increase of $|r|$ (the actual infinity). One can see that it decreases, and upon a closer inspection, it seems like the imaginary part of the modes tends to a constant while the real part tends to zero. We could not reach that probable limit up to $|r|=1000$, where the calculations become computationally too expensive. One can see hints of such a limit also on Fig. \ref{Fig5} b), where a complex plot of 5 consecutive modes in the interval for $|r|=110..250$ is plotted. On it $|r|=110$ is to the right and there is a hint of convergence for the different modes. For the moment, however, such a limit cannot be definitely confirmed numerically.
 \item Stability with respect to the argument of the radial variable

 The study of the dependence of the modes on $arg(r)$ using the $\epsilon$-method can be seen on Fig. \ref{Fig6}. On it, we have varied $\epsilon=0..\pm1$ for one mode with $l=1$ (0.044511+.297566i) and one -- with $l=2$ (0.023201+0.110030i). The interval for $\epsilon$ means that the argument is moved in a complete circle in the unit-sphere. As can be seen from the figure, the complex plot of those modes also draws a circle. One can see also the boundary conditions in this case. As mentioned in \cite{arxiv2014}, the spurious modes are obtained from both equations ($R_1(r)$ and $R_2(r)$) for the same value of argument -- the left branch for $arg(r)=1/2\pi$, the right -- for $\arg(r)=3/2\pi$. On the plots we have used only points calculated from $R_2(r)$ but the values obtained from $R_1(r)$ coincide with them to a precision of $10^{-6}$.

 \item Stability with respect to rotation

 On Fig. \ref{Fig5} c) we plotted the behavior of one spurious mode with respect to an increase of the rotation from $a=0..M$. One can see the characteristic for the confluent Heun functions loops, which occur also for the QNM modes. Since those modes evolve with rotation all the way to the naked singularity regime, it is important to be able to distinguish them from the QNM/QBM modes by checking the $|r|$ dependence of the resulting frequencies and the boundary conditions. This problem is the most serious for the QNM mode with $n=6$ which lies near a spurious mode, (see Fig. \ref{Fig1}), but considering how densely spaced is the spurious modes spectrum, it could occur also near other modes (for $a\in [0,M]$).
 \end{enumerate}

It is important to note that those modes are roots not only to the radial equation, but to the spectral system. They can be seen to propagate for $a>0$ and even in the naked singularity case -- for $a>M$ while the QNM/QBM spectra do disappear for $a>M$ or at least for the moment cannot be located by the root-finding algorithm. Because of their persistence in the numerical results, the only way to recognize the spurious spectra is through the radial stability checks we performed.

\section{Discussions}
It is important when discussing the numerical stability of a spectrum to make the difference between the numerical stability of the frequencies, the stability of the obtained spectrum and the stability of the spacetime with the given metric. Above, we emphasized that we discussed the numerical stability of the results as an essential test of how trustable they are. When dealing with numerical solutions, we always work with certain precision and in a certain numerical domain, where the numerical algorithm is applicable. The latter depends on both the theory of the function in question and the numerical routines evaluating it.

In our case, the function we are working with is the confluent Heun function. Because of the irregularity of the singular point at infinity, there are no currently known series expansions working at infinity, and the available asymptotics of the solutions of the confluent Heun equation is not properly developed, too. Thus, one has limited options when dealing with infinity. The method of the continued fraction is one possible way to deal with the problem. With it, one moves the singularities, so that: $0 \to \infty, 1\to 0, \infty \to 1$. In this way, the series is convergent from $z=0$ to the next nearest singularity $z=1$ which corresponds to infinity. With this method, one obtains a continued fractions for the coefficients of the recurrence, from which the $r-$variable is missing. The limits of applicability of this method are only partially known -- for example, it does not work for purely imaginary modes. What is more important, this method does not address the fact that there are 16 classes of solutions of the confluent Heun type valid in different parts of the complex plane. There is also the problem with the branch cuts spectra needed for the full description of the perturbation (\cite{BC3, BC4}). Finally, observations of the effect of the Stokes and Anti-Stokes lines were noted in works utilizing the phase-amplitude method by Andersson \cite{Andersson}. All those details point to the need of proper development of the theory of the confluent Heun equation and its solutions, in order to have clarity about the area of applicability of the different methods and their physical interpretation. In our case, to test the numerical stability of our results, we compare the output of the 3 methods of evaluation of the confluent Heun function currently implemented in Maple -- direct integration, Taylor expansion and continuation of the series expansion around $z=1$, and then we check if small deviations in the parameters lead to significant changes in the frequencies. We also plot the Stokes and Anti-Stokes contours, as well as a 3rd complex plot of the function around a root for $a=0$. All those tests point to numerical stability of the QNM and QBM spectra and to instability of the third one.

It is important to note that the QNM problem in black holes is an analogue to the one of finding the poles of the S-matrix of the Schrodinger equation. In Quantum Mechanics, those poles depending on the boundary conditions may correspond to bound states, virtual states or resonant states. All of them correspond to different boundary conditions in which one may deal with solutions with exponentially decreasing or increasing asymptotics (bound states or resonances). In astrophysics, because of the existence of the horizon on one of the boundaries of the S-matrix and, furthermore, the fact that all the frequencies are complex, one cannot apply directly the same terminology and methods to the problem. For this reason, we work with quasinormal modes and quasibound modes, even though the bound states in Quantum Mechanics are purely imaginary modes. In quantum mechanics, the branch cut of the S-matrix also plays an important and well-known role. In the theory of QNM we have currently a very limited number of works on this approach (\cite{BC1, BC2, BC3},\\ \cite{BC4}).

Apart from the numerical stability, one also may discuss the physical stability of the spectrum. Here with ''spectrum``, we refer to the solutions of the spectral problem defined by the exact solutions of the differential equations and the boundary conditions we impose on them. As known from the theory of the differential equations, different boundary conditions describe different physical objects, even though they are governed by the same differential equation. In this context, we can divide the different spectra into stable and unstable. For example, the spectrum obtained from black hole boundary conditions in our work has been found to be physically stable, i.e. corresponding to perturbations which damp with time. On the other hand, the spectrum corresponding to the quasibound modes is clearly physically unstable, because the perturbations grow with time. There could be an other explanation, however. As mentioned in \cite{arxiv2014}, the ranges of the Boyer-Lindquist coordinates for the Kerr metric are: $0\le r \le \infty, 0\le \theta \le \pi, 0 \le \phi \le 2\pi, -\infty <t<\infty$. In the case of the standard substitution $\Psi=e^{i(\omega t+m\phi)}S(\theta)R(r)$, the perturbation will damp with time for $\Im(\omega)>0$ and $t>0$. For $t<0$ the perturbation will grow exponentially for the same condition. For the QBM spectra, the situation reverses and for $\Im(\omega)<0$ the perturbations will damp with time for $t<0$ and grow exponentially for $t>0$. Considering the full range of the variables is not unusual. For example in \cite{neg_mass}, the standard extension $-\infty < r \le 0, M>0$, equivalent to $r\ge 0, M<0$ was already considered. Also, it is known that the metric is invariant under the changes $t\to-t, \phi \to -\phi$, meaning that the inversion of time can be considered as inversion of the direction of rotation. Therefore, both QNM and QBM are valid frequencies in the whole range for the variables of the problem.

Finally, the question of the stability of the physical object described by the metric is more complicated and has a long history (for example \cite{stability1, stability2, stability3, stability4}).  In many of those studies, linear perturbation theory was used with certain ambiguity in the definitions. When one speaks of the stability of the Schwarzschild black hole, one implies that the boundary conditions will be those of a black hole (BHBC). The Schwarzschild metric, however, just like the Kerr metric, in linear perturbation theory, allows one to impose other types of boundary conditions -- both the inverse BH conditions which we call QBBC or the total-transmission modes boundary conditions. In those cases, however, one deviates from the classical description of a black hole since the waves are not entering the horizon, but moving away from it. This problem becomes even more complicated when we work with the Kerr metric. The Kerr metric according to the theory has 2 horizons ($r_\pm$) and one physical singularity ($r=0$), which is singularity of the scalar invariants of the metric. This definition usually ignores the differential invariants where the horizons are also visible. What is more important, the singularities visible for the differential equation when $a\neq 0$ are only the two horizons and infinity. Because of this, following the theory of ordinary differential equations, one needs to impose the boundary conditions on two of those singularities in order to specify the singular spectral problem. Usually, one regards the inner horizon as physically not interesting and imposes the boundary conditions on the outer event horizon and infinity. But there have been studies in which one imposes the boundary conditions on two horizons, which have shown that the region between the two horizons is {\em unstable} with respect to linear perturbations. This poses the question of the physical interpretation of the different regions of the Kerr metric and their analytical extensions. In any case, there are 3 different types of boundary conditions -- BHBC, QBBC and TTM -- which correspond to different physical situations occurring on the background of the Kerr metric and their study has its theoretical value.

\section{Conclusion}
The quasinormal modes are a well-known and seriously studied problem in gravitational physics. They have been extensively used in different studies, but one significant problem has remained and it is related to the theory of the confluent Heun functions and their roots. In series of articles, we have studied different physical configurations using the confluent Heun functions in Maple and we demonstrated that this method can lead to interesting scientific results, however, this novel approach also requires the development of new numerical tools and methods that are able to deal with the complexity of the Heun differential equations and their solutions.

In this article, by using the exact solutions of the TRE and TAE, we were able to impose different boundary conditions on them, describing QNM and QBM. By checking the numerical stability of the so obtained spectra in both $|r|$ and $\arg(r)$, we were able to sift out the physical solutions from the spurious ones, with clear theoretical justification. This proves that the new approach indeed leads to a better understanding of the problem in question and also points to the huge potential which any improvement in the theory of the Heun functions offers to physics.

\section{Acknowledgements}
The authors would like to thank Dr. Edgardo Cheb-Terrab for useful discussions of the algorithms evaluating the Heun functions in \textsc{maple}, and for continuing the
improvement of those algorithms in the latest versions of \textsc{maple}.

The authors would like to thank D. Sc. Svetlana Pacheva for critical reading of the manuscript and the constructive comments.

This article was supported by the Foundation "Theoretical and
Computational Physics and Astrophysics", by the Bulgarian National Scientific Fund
under contracts DO-1-872, DO-1-895, DO-02-136, and Sofia University Scientific Fund, contract 185/26.04.2010, Grants of the Bulgarian Nuclear
Regulatory Agency for 2013, 2014 and 2015.

\section{Author Contributions}
P.F. posed the problem of the evaluation of the EM QNMs of rotating BHs as a continuation of previous studies of the applications of the confluent Heun functions in astrophysics. He proposed the epsilon method and supervised the project.

D.S. is responsible for the numerical results, their analysis and the plots and tables presented here.

Both the authors discussed the results at all stages. The manuscript was prepared by D.S. and edited by P.F..
\appendix{}
\section{Tables of the obtained EM QNMs}
Table \ref{table_n0_om}  presents some of the values obtained for the EM QNM, converted to physical units using the relations:
$$\omega^{phys}=\Re(\omega)\frac{c^3}{2\pi\,G\,M}, \tau^{phys}=\frac{1}{\Im(\omega)}\frac{GM}{c^3}.$$

Note that in those formulas a factor of $2$ is missing because the EM QNMs were obtained for $M_{KBH}=1/2$ and not for $M_{KBH}=1$. Then if $M$ is the mass of the object in physical units (we use  $M=10M_{\odot}$), $M_{\odot}$ -- the mass of the Sun ($M_{\odot}=1.98892\,10^{30}[kg]$) and  $G=6.673\,10^{-11}[\frac{m^3}{kg\,s^2}], c=2.99792458\,10^8[m/s]$, one obtains $$\omega^{phys}\approx \frac{32310}{M/M_{\odot}}\Re(\omega)[Hz],\tau^{phys}\approx \frac{0.4925\,10^{-5}\,M/M_{\odot}}{\Im(\omega)}[s].$$

%The frequencies and the damping times in the table are calculated for.
\begin{table*}[!h]	
\footnotesize
\begin{tabular}{|l | l  | l | l | l | l | l|}
\hline  \multicolumn{7}{|c|}{$n=0$}\\
\hline
& \multicolumn{2}{|c|}{$m=0$} & \multicolumn{2}{|c|}{$m=-1$} & \multicolumn{2}{|c|}{$m=1$}\\
\hline \multirow{1}{*}{$a/M$}  & $\omega^{phys}_{m=0}$[Hz] & $\tau^{phys}_{m=0}$[ms] &  $\omega^{phys}_{m=-1}$[Hz] & $\tau^{phys}_{m=-1}$[ms] &  $\omega^{phys}_{m=1}$[Hz] & $\tau^{phys}_{m=1}$[ms]\\
\hline
0&    802.1512449166& 0.5325890917& 802.1512449166& 0.5325890917&  802.1512449167& 0.5325890917\\
0.2& 804.9393652797& 0.5343356142& 849.8315682698& 0.5388677452&   763.6902591869& 0.5299212818\\
0.6&  829.2637578502& 0.5526525743&   996.9258848852& 0.5772488810& 704.6451920585& 0.5313970112\\
0.98& 884.6086875757& 0.6427140687& 1445.8841670353& 1.2178343064&  661.8628389523& 0.5373275077\\
\hline
\multicolumn{7}{|c|}{$n=3$}\\
\hline
& \multicolumn{2}{|c|}{$m=0$} & \multicolumn{2}{|c|}{$m=-1$} & \multicolumn{2}{|c|}{$m=1$}\\
\hline \multirow{1}{*}{$a/M$} & $\omega^{phys}_{m=0}$[Hz] & $\tau^{phys}_{m=0}$[ms] &  $\omega^{phys}_{m=-1}$[Hz] & $\tau^{phys}_{m=-1}$[ms] &  $\omega^{phys}_{m=1}$[Hz] & $\tau^{phys}_{m=1}$[ms]\\
\hline
0&    472.3043572607 & 0.0638131626 & 472.3043572599 & 0.0638131626 & 472.3043572609 & 0.0638131626 \\
0.2 & 479.7880705182 & 0.0641624606 & 549.5382706431 & 0.0658306413 & 415.5213067645 & 0.0624052214 \\
0.6 & 530.3546588895 & 0.0677548111 & 792.4033820584 & 0.0741706889 & -- & -- \\
0.98 & 408.1590806664 & 0.0756272782 & -- & -- & -- & -- \\
\hline
\end{tabular}
\caption{Table of the frequencies, $\omega^{phys}$, in Hz , the damping times, $\tau^{phys}$, in miliseconds for $n=0,3$, $l=1$ for some chosen values of the rotational parameter $a$. Here $M= 10 M_\odot$.}
\label{table_n0_om}
\end{table*}
\vspace{-1.5cm}
\begin{table*}[!h]	
\footnotesize
\begin{tabular}{|l | l  | l | l |}
\hline  \multicolumn{4}{|c|}{$n=0$}\\
\hline
& \multicolumn{1}{|c|}{$m=0$} & \multicolumn{1}{|c|}{$m=-1$} & \multicolumn{1}{|c|}{$m=1$}\\
\hline \multirow{1}{*}{$a/M$}  & $E_{m=0}$ &  $E_{m=-1}$ & $E_{m=1}$ \\
\hline
0& 2.0000000000 + $9.4 10^{-63}i$&  2.0000000000 + $4.82 10^{-63}i$&   2.0000000000 + $4.83 10^{-60}i$\\
0.2& $1.9991429248 - 0.0007350608i$& $1.9460508578 + 0.0193509134i$&       $2.0462372214 + 0.0176327659i$\\
0.6& $1.9916552351 - 0.0066017264i$&       $1.7970919424 + 0.0620215234i$&       $2.1232126190 + 0.0478089014i$\\
0.98&$1.9734060595 - 0.0162271167i$&       $1.4500680177 + 0.0605778034i$&       $2.1833506585 + 0.0705481152i$\\
\hline
\multicolumn{4}{|c|}{$n=3$}\\
\hline
& \multicolumn{1}{|c|}{$m=0$} & \multicolumn{1}{|c|}{$m=-1$} & \multicolumn{1}{|c|}{$m=1$}\\
\hline \multirow{1}{*}{$a/M$}  & $E_{m=0}$ &  $E_{m=-1}$ &  $E_{m=1}$\\
\hline
0& 2.0000000000+$9.44 10^{-31}$i& 2.0000000000+4.82 $10^{-30}$i&  2.0000000000+4.83 $10^{-30}i$\\
0.2&$2.0090686631-0.0036398098i$& $1.9778301381+.1550332409i$&  $2.0389149509+.1531306042i$\\
0.6&$2.0717320628-0.0337575846i$& $1.9332832523+.4598448835i$& -- \\
0.98&$2.1483751533-0.0584019594i$&$1.6286222388+.6109228338i$&--\\
\hline
\end{tabular}
\caption{Table of the separation parameter $E$ for $n=0,3$, $l=1$ for some chosen values of the rotational parameter $a$. Here $M= 10 M_\odot$.}
\vspace{-0.5cm}
\label{table_n0_E}
\end{table*}

\end{document}